%% file: CesrConvWiggler-JINST.tex
\pdfoutput=1
\documentclass{JINST}

\usepackage{amssymb}
\usepackage{lineno}
\usepackage{graphicx}
\usepackage{amssymb}
\usepackage{amsmath}
\usepackage{epstopdf}
\usepackage{relsize}

\def\cesrta{{C{\smaller[2]ESR}TA}}
\include{definitions}

\title{The Conversion of CESR to Operate as the Test Accelerator, CesrTA, Part 4: Superconducting Wiggler Diagnostics}

\author{M.G.Billing, S.Greenwald, X. Liu, Y.Li, D.Sabol, E.N.Smith, C.R.Strohman,\
Cornell Laboratory for Accelerator-based Sciences and Education,\
Cornell University, \\161 Synchrotron Dr., Ithaca, NY, 14850, U.S.A.}
\author{M.A.Palmer,\
Collider-Accelerator Department, Brookhaven National Laboratory,\\
Bldg. 911B, P.O. Box 5000, Upton, NY 11973-5000, U.S.A.}
\author{D.V.Munson,\
Lawrence Berkeley National Laboratory,\\
1 Cyclotron Rd., Berkeley, CA 94270, U.S.A.}
\author{Y. Suetsugu,\
High Energy Accelerator Research Organization (KEK),\\
1-1 Oho, Tsukuba, Ibaraki Prefecture 305-0801, Japan}

\abstract{ Cornell's electron/positron storage ring (CESR) was
modified over a series of accelerator shutdowns beginning in May
2008, which substantially improves its capability for research and
development for particle accelerators.  CESR's energy span from 1.8
to 5.6 GeV with both electrons and positrons makes it 
appropriate for the study of a wide spectrum of
accelerator physics issues and instrumentation related to present
light sources and future lepton damping rings. Additionally a number
of these are also relevant for the beam physics of proton
accelerators.  This paper, the last in a series of four,
describes the vacuum system
modifications of the superconducting wigglers to accommodate the
diagnostic instrumentation for the study of electron cloud (EC) behavior within
wigglers.  Earlier papers provided an overview of the accelerator physics program, the 
general modifications of CESR, the modifications of the vacuum
system necessary for the conversion of CESR to the test accelerator,
{\cesrta}, enhanced to study such subjects as low emittance tuning
methods, EC effects, intra-beam scattering, fast
ion instabilities as well as general improvements to beam
instrumentation.  While the initial studies of {\cesrta} focussed on
questions related to the International Linear Collider damping
ring design, CESR is a very versatile storage ring, capable of
studying a wide range of accelerator physics and instrumentation
questions.}


\begin{document}

\input{overview}
\input{vacuum_system}

\section{Summary}

The installation of a significant number of vacuum chambers with their associated diagnostics were required for the modification of the storage ring CESR to support the creation of {\cesrta}, a test accelerator configured to study accelerator beam physics issues for a wide range of accelerator effects and to develop instrumentation related to present light sources and future lepton damping rings.  This paper has presented
many of the details for the modifications of the superconducting wigglers to accommodate the
diagnostic instrumentation for the study of EC behavior within
wigglers. This paper is a useful reference for any future analysis of the {\cesrta}  measurements of electron clouds  in superconducting wigglers.  As part of the larger {\cesrta} program, CESR's vacuum system and instrumentation have been optimized for the study of low emittance tuning methods, electron cloud effects, intra-beam scattering, fast ion instabilities as well as the development and improvement of beam diagnostics.

\label{sec:summary}

\section*{Acknowledgements}

The authors would like to acknowledge the many contributions that have helped make the {\cesrta} research program a success. It would not have occurred without the support of the International Linear Collider Global Design Effort led by Barry Barish. Furthermore, our colleagues in the electron cloud research community have provided countless hours of useful discussion and have been uniformly supportive of our research goals.

We would also like to thank the technical and research staff at Cornell{'}s Laboratory for Accelerator ScienceS and Education (CLASSE) for their efforts in maintaining and upgrading CESR for Test Accelerator operations. Specifically we would like to acknowledge the contributions of a few individuals during this portion of the project:  Margie Carrier performed the majority of RFA (in-vacuum) electric assembly, which was extremely complex and tedious.  James Sears and Brian Clasby made significant contributions to the e-beam welding of the final RFA cover.  They had to add an extension to the Newman Laboratory electron-beam welder chamber, 
re-program the e-beam welding head, work out detailed welding parameters, etc.  Ken Powers assisted with the  removal of old SCW beampipes and preparations of the SCEs for the RFA beampipes.  Mike Palmer produced excellent welding results for the final assembling of the RFA beampipes into the SCWs. 

Finally, the authors would like to acknowledge the funding agencies that helped support the program. The U.S. National Science Foundation and Department of Energy implemented a joint agreement to fund the {\cesrta} effort under contracts PHY-0724867 and DE-FC02-08ER41538, respectively. Further program support was provided by the Japan/US Cooperation Program. Finally, the beam dynamics simulations utilized the resources off the National Energy Research Scientific Computing Center (NERSC) which is supported by the Office of Science in the U.S. Department of Energy under contract DE-AC02-05CH11231.





\bibliographystyle{amsplain}
\bibliography{Bibliography/CesrTA}







\end{document}

%% file: definitions.tex
\def\cesrta{{C{\smaller[2]ESR}TA}}

\newcommand{\beq} {\begin{equation}}
\newcommand{\eeq} {\end{equation}}

%% file: overview.tex
\section{Overview of CESR Modifications}
\label{sec:cesr_conversion.overview}

The conversion of CESR to permit the execution of the \cesrta\ program \cite{JINST10:P07012} required several extensive modifications.  These included significant reconfiguring of CESR's accelerator optics by removing the CLEO high energy physics (HEP) detector and its interaction region, moving six superconducting wigglers and reconfiguring the L3 straight section\cite{JINST10:P07012}.  There were also major vacuum system modifications to accommodate the changes in layout of the storage ring guide-field elements, to add electron cloud diagnostics and to prepare regions of the storage ring to accept beam pipes for the direct study of the electron cloud\cite{JINST10:P07013}.  A large variety of instrumentation was also developed to support new electron cloud diagnostics, to increase the capabilities of the beam stabilizing feedback systems and the beam position monitoring system, to develop new X-ray beam size diagnostics and to increase the ability for studying beam instabilities\cite{JINST11:P04025}.  This conversion process included the development of special instrumentation for the study of EC within superconducting wigglers and these are described in the following section.  The entire conversion occurred in down periods over several years, during which the operation of the Cornell High Energy Synchrotron Source (CHESS) was maintained for synchrotron x-ray users.

\subsection{Motivation for CESR Modifications}
\label{ssec:cesr_conversion.vac_system.overview}
\begin{figure}
    \centering
    \includegraphics[width=0.75\textwidth]{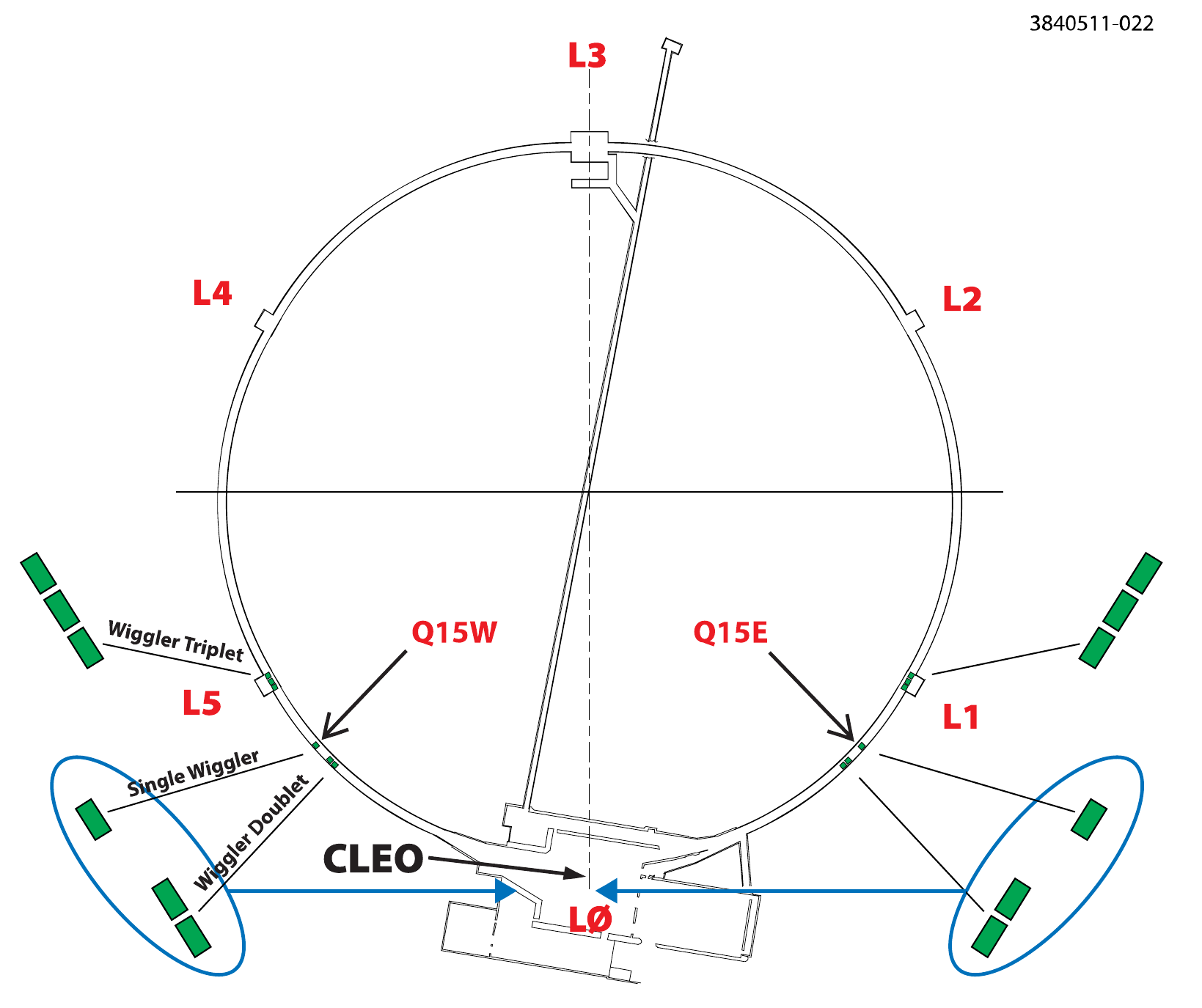}
    \caption[{\cesrta} Vacuum System]{The reconfiguration of the CESR vacuum system provided space in two long experimental regions in L0 and L3, and two flexible short regions near Q15W and Q15E. Hardware for electron cloud studies was installed in these regions\cite{JINST10:P07013}. The west side of the L0 experimental region housed the retarding field analyzer- (RFA-)equiped SCWs. \label{fig:cesr_conversion:vac_fig1}}
\end{figure}

  CESR's vacuum system is an essential part of its accelerator beam transport system, which is capable of storing total electron and positron beam currents up to 500~mA (or single beam up to 250 mA) at a beam energy of 5.3~GeV.  As shown in Figure~\ref{fig:cesr_conversion:vac_fig1} the CESR vacuum system with a total length of 768.44~m consists primarily of bending chambers in the arcs, two long straight sections, called $L0$ (18.01~m in length) and $L3$ (17.94~m in length), and four medium length straights (called $L1$~and~$L5,$ both 8.39~m in length and $L2$~and~$L4,$ both 7.29~m in length).

   In the CESR-c/CLEO-c era 12 home-built superconducting wigglers (SCWs) were installed in CESR\cite{PAC03:WPPE037}, with two triplets at $L1$ and $L5$, two doublets and two singlet SCWs in the arcs (see Figure~\ref{fig:cesr_conversion:vac_fig1}).  This complement of SCWs plays an important role in the {\cesrta} program\cite{JINST10:P07012, JINST10:P07013, JINST11:P04025}.  These SCWs additionally represent a good model for the wigglers planned for the International Linear Collider (ILC) damping rings\cite{IPAC12:TUPPR065}.
At the conclusion of CLEO-c HEP program in March 2008, staged modifications were carried out to convert CESR into the test accelerator {\cesrta} \cite{JINST10:P07012, JINST10:P07013, JINST11:P04025}.  The motivation for the vacuum system modification was to support the physics programs of {\cesrta}, including: (1)~Ultra-low emittance lattice design, tuning and associated beam instrumentations, and (2)~Electron cloud studies, including the development and verification of suppression techniques for ECs.

During the reconfiguration of L0, including the removal of the central part of the CLEO HEP Detector and the IR vacuum chambers, 6 SCWs [4 CESR-c and two \cesrta\ SCWs, equipped with retarding field analyzers (RFA)] were installed along with many new chambers containing additional EC diagnostics, such as RFAs and beam position monitor (BPM)/TE-wave buttons.

%% file: vacuum_system.tex


\section[Wiggler Chambers]{Wiggler Chambers}
\subsection{Design Considerations for Wiggler RFA Structure}

Superconducting wigglers (SCWs) were designed and fabricated for the CESR-c/CLEO-c program.  The CESR-c wigglers are 8-pole super-ferric magnets with the main period of 40~cm and trimming end poles.  The SCWs provide a very uniform transverse field up to 2.1~T, closely matching the ILC DR wiggler requirements.  Therefore, these SCWs are ideal test vehicles for the study of EC growth and suppression in the wiggler field.  With this motivation a thin-style RFA design was developed for the wiggler beam pipe and a set of diagnostic chambers was constructed by a collaboration including Cornell, KEK, LBNL and SLAC.

\begin{figure}
	\centering
	\includegraphics[width=0.9\textwidth]{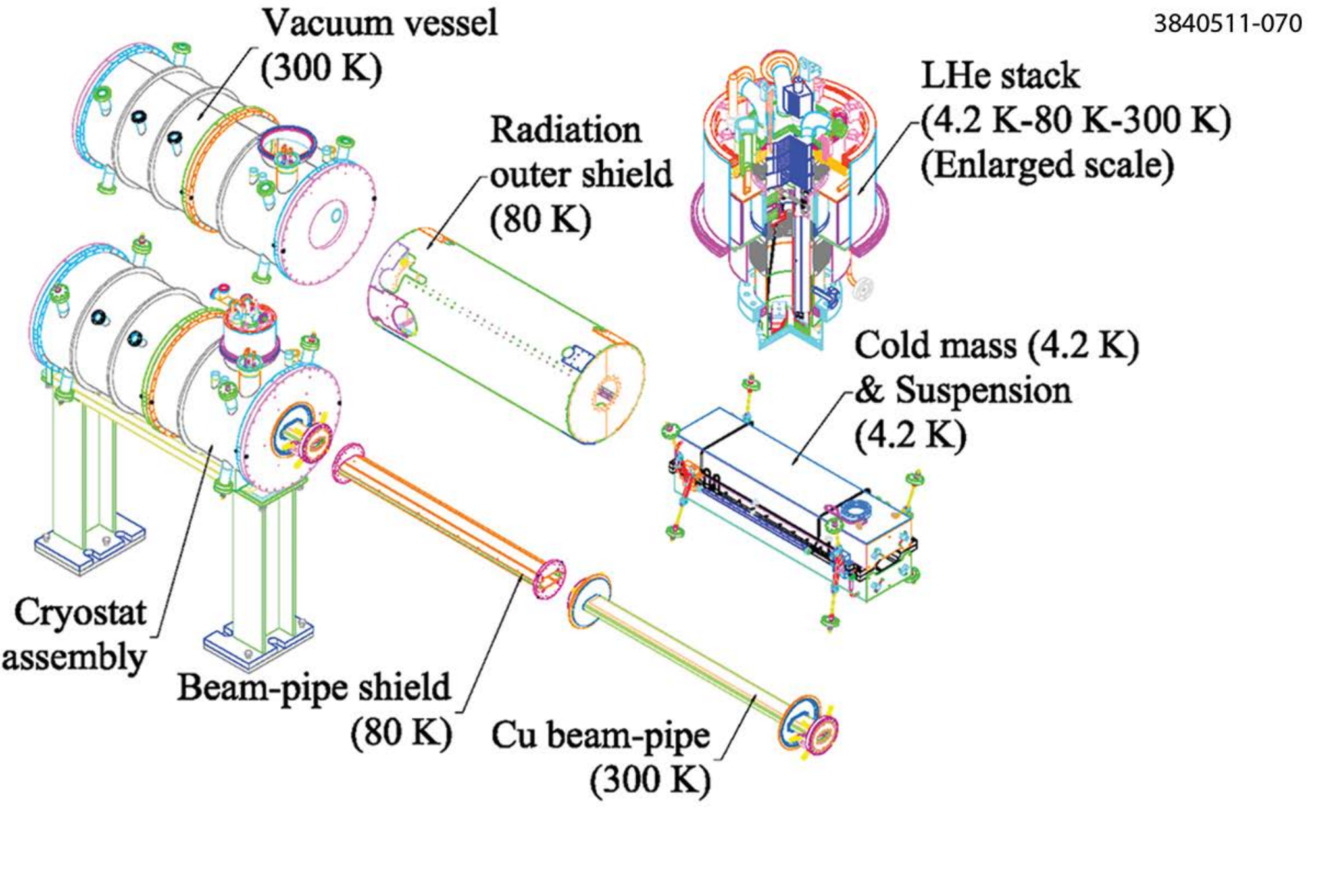}
	\caption{Exploded view of the structure of a CESR-c superconducting wiggler assembly. \label{fig:SCW_CESR_c}}	
\end{figure}

Figure~\ref{fig:SCW_CESR_c} illustrates the structure of the SCW. The details regarding the design of the CESR-c SCWs can be found in \cite{PAC03:WPPE037}.  In the original SCW production, the beam pipe was assembled into the magnet's cold-mass and the cold-mass with the beam pipe was in turn inserted in the isolation vacuum vessel.  Sixteen SCWs were constructed for the CESR-c program on a specially designed SCW assembly line in a dedicated facility that no longer exists.  Therefore, the following design constraints were imposed on the RFA beam pipe design and construction.

\begin{itemize}
	\item The RFA beam pipe must fit inside the space in the cold-mass thermal shield (see Figure~\ref{fig:SCW_CESR_c}) while still providing sufficient beam aperture.
	\item The removal of the existing beam pipe and the insertion of the RFA beam pipe must be accomplished without disassembling or disturbing the wiggler magnet structure.
	\item The RFA beam pipe (including all material used) must be UHV-compatible. 
\end{itemize} 

\begin{figure}
	\centering
	\includegraphics[width=0.6\textwidth, angle=-90, width=6.0in]{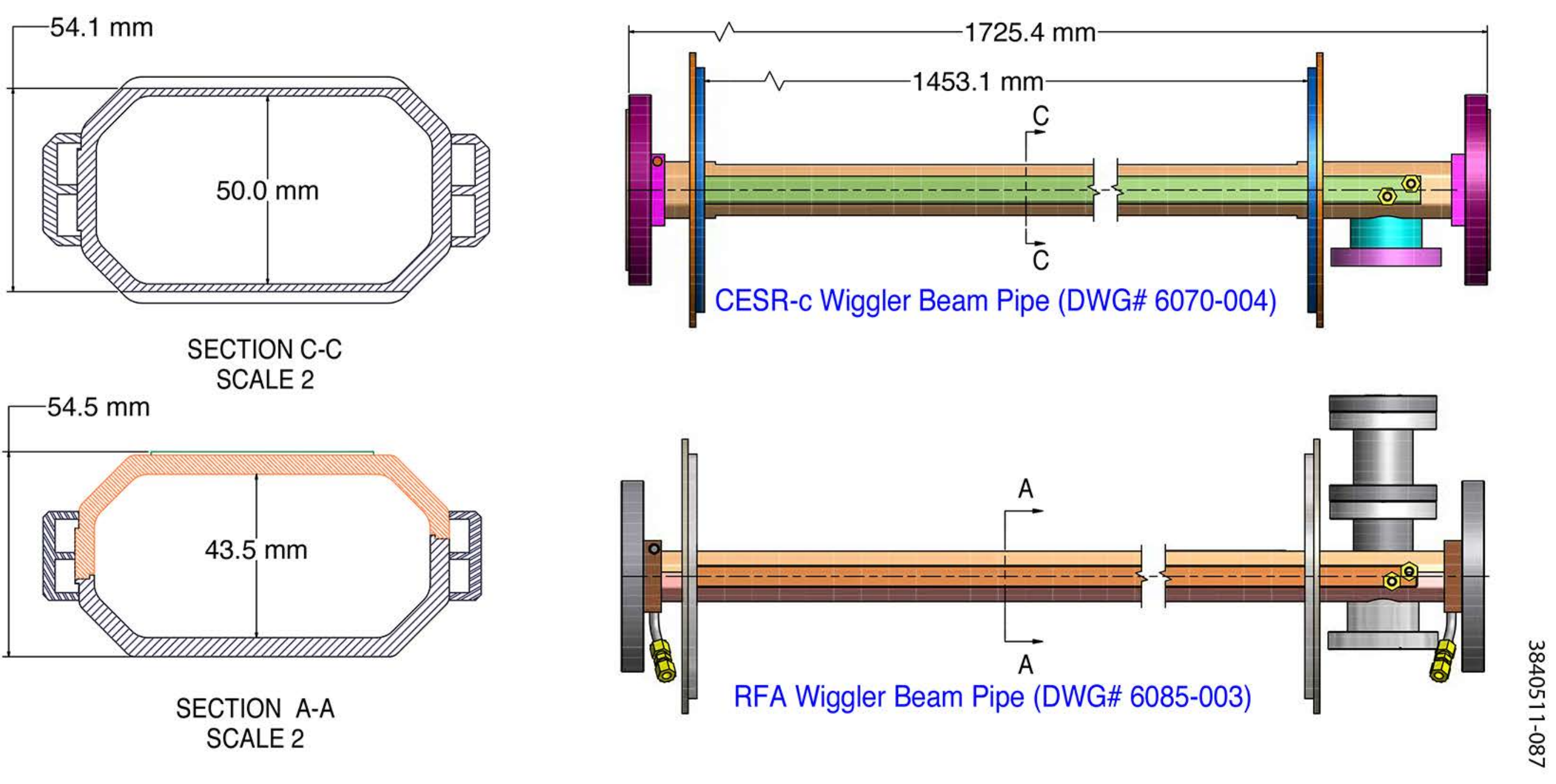}
	\caption{Cross section comparison between a `standard' CESR-c SCW beam pipe and a {\cesrta} RFA-equipped SCW beam pipe. \label{fig:SCW_beampipe_Comparison}}
\end{figure}

	With these requirements the vertical aperture of the RFA beam pipe was reduced from the standard 50~mm to 43.5~mm to provide space for the RFAs, as shown in Figure~\ref{fig:SCW_beampipe_Comparison}.  The copper extrusions were split into top and bottom halves.  After trimming the vertical edges, the two halves were re-joined (via electron beam welds) to form the RFA beam pipe.  

Over the first three years, four SCWs equipped with RFA beam pipes, were constructed and installed for the \cesrta\ program \cite{JINST10:P07012, JINST10:P07013, JINST11:P04025}.  The next five sections describe this process for implementing RFAs in the superconducting wiggler magnets in some detail.  This is to provide adequate documentation to support the integrity of the instrumentation and to address concerns, which might call the measurements into question.  They also provide guidance for any future developments of wiggler-based RFAs.  Lastly, these sections allow enough details for future simulations of EC development in these SCWs.

Of the first two units, one has a bare copper beam pipe and the other has a copper beam pipe coated with a TiN thin film (prepared by the SLAC team.)  In the third and fourth units, a grooved plate and an EC clearing electrode were implemented on the bottom half of the beam pipe.  Details for these four assemblies are found in the following sections.

\subsection{Overview of RFA Construction}
		
	As detailed later in this section, the top half beam pipe houses the RFA assemblies, while a variety of EC suppression techniques were implemented on the bottom half beam pipe.  The experimental program for the wiggler vacuum chamber mitigation studies occurred as summarized below.
	
\begin{itemize}
	\item  A bare copper chamber installed for a reference SCW measurements in October 2008.
	\item  Copper chamber coated with TiN also installed in October 2008.
	\item  Grooved SCW chamber placed in CESR in July 2009 and operated in experimental runs (July 2009 to September 2009, and November 2009 to December 2009) and two CHESS runs (September 2009 to November 2009, and January 2010 to March 2010.)
	\item  In the April 2010 CESR shutdown, the preceding SCW assembly was replaced with a SCW vacuum chamber fitted with RFA beam pipe, having an EC clearing electrode.
	\item  The grooved RFA SCW assembly was coated with TiN and then reinstalled in the L0 EC experimental region in January 2011.
\end{itemize}

	Simulations \cite{PAC07:TUPAS067} predicted distinct longitudinal and transverse EC density distributions in wiggler beam pipes.  The design goal of the wiggler RFA beam pipe is to place RFAs at strategic longitudinal locations in the wiggler magnetic field to measure the corresponding transverse EC density distributions.

\begin{figure}
	\centering
	\includegraphics[angle=-90, width=0.85\textwidth]{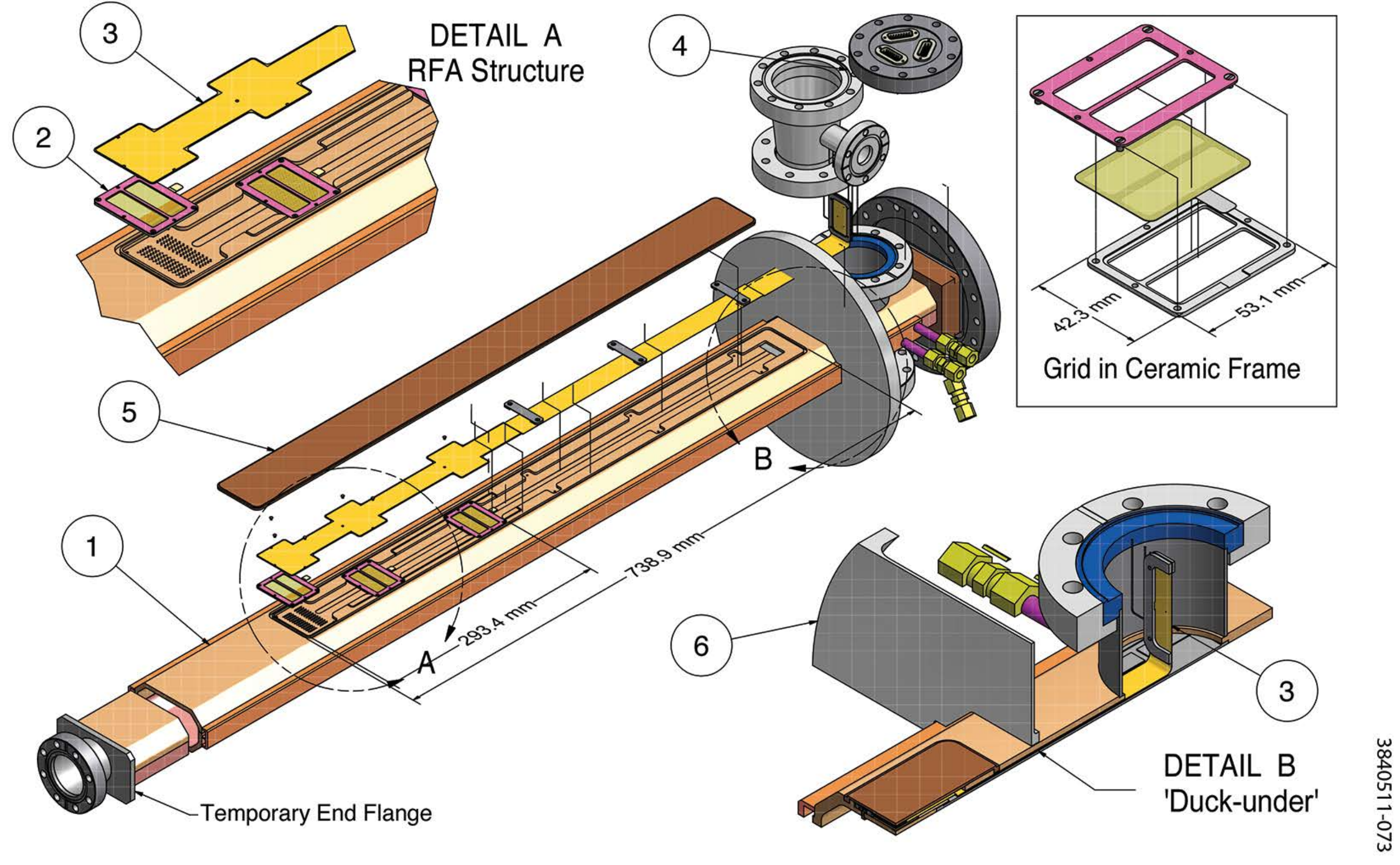}
	\caption[Exploded View of a SCW RFA beam pipe Assembly]{Exploded View of a SCW RFA beam pipe Assembly. The key components are: (1) beam pipe top half, housing the RFAs; (2) RFA grids (see upper right inset); (3) RFA collector on a flexible printed circuit board; (4) RFA connection port; (5) RFA vacuum cover; (6) flexible disk, sealing the beam pipe-insulation vacuum interface and allowing for thermal expansion of the beam pipe.  The `duck-under' channel, through which the kapton flexible circuit is fed after all heavy through-welding is complete, is shown in detail B.
\label{fig:SCW_RFA_structure}}
\end{figure}

\begin{figure}
	\centering
	\includegraphics[width=0.85\textwidth]{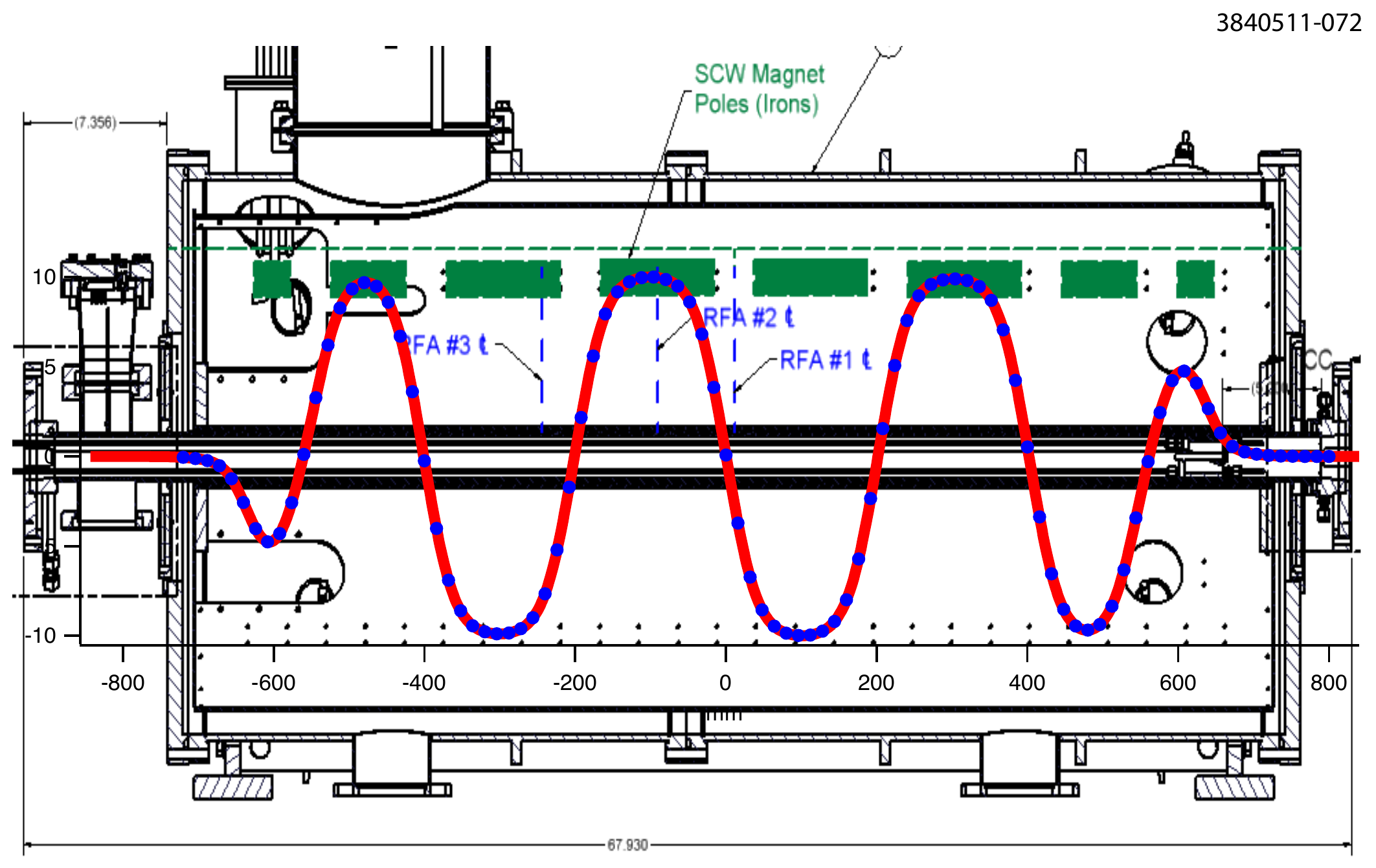}
	\caption{Three RFAs are built into each SCW RFA beam pipe.  A plot of the vertical B-field along the wiggler (red line with blue dots) is superimposed on the drawing of the wiggler.  The centers of the RFAs are located at three strategic B-field locations, as shown. \label{fig:SCW_RFA_locations}}
\end{figure}

	The RFA beam pipe design 
is illustrated in an exploded view in Figure~\ref{fig:SCW_RFA_structure}. The RFA beam pipe is constructed from the top and bottom halves of the same extrusions as used for original SCW beampipes.  Along the longitudinal direction of the beam pipe, three pockets were machined into the top half of the beam pipe to form housings for the three RFAs.  The three RFAs' locations together with the vertical field map of the SCW are depicted in Figure~\ref{fig:SCW_RFA_locations}.  After installation in the wiggler magnet, the RFA \#1 is at boundary between two center poles (zero vertical magnetic field), \#2 at center of a pole (maximum vertical magnetic field) and \#3 at `edge' of a pole (longitudinal B-field).

\begin{figure}
	\centering
	\includegraphics[width=0.75\textwidth]{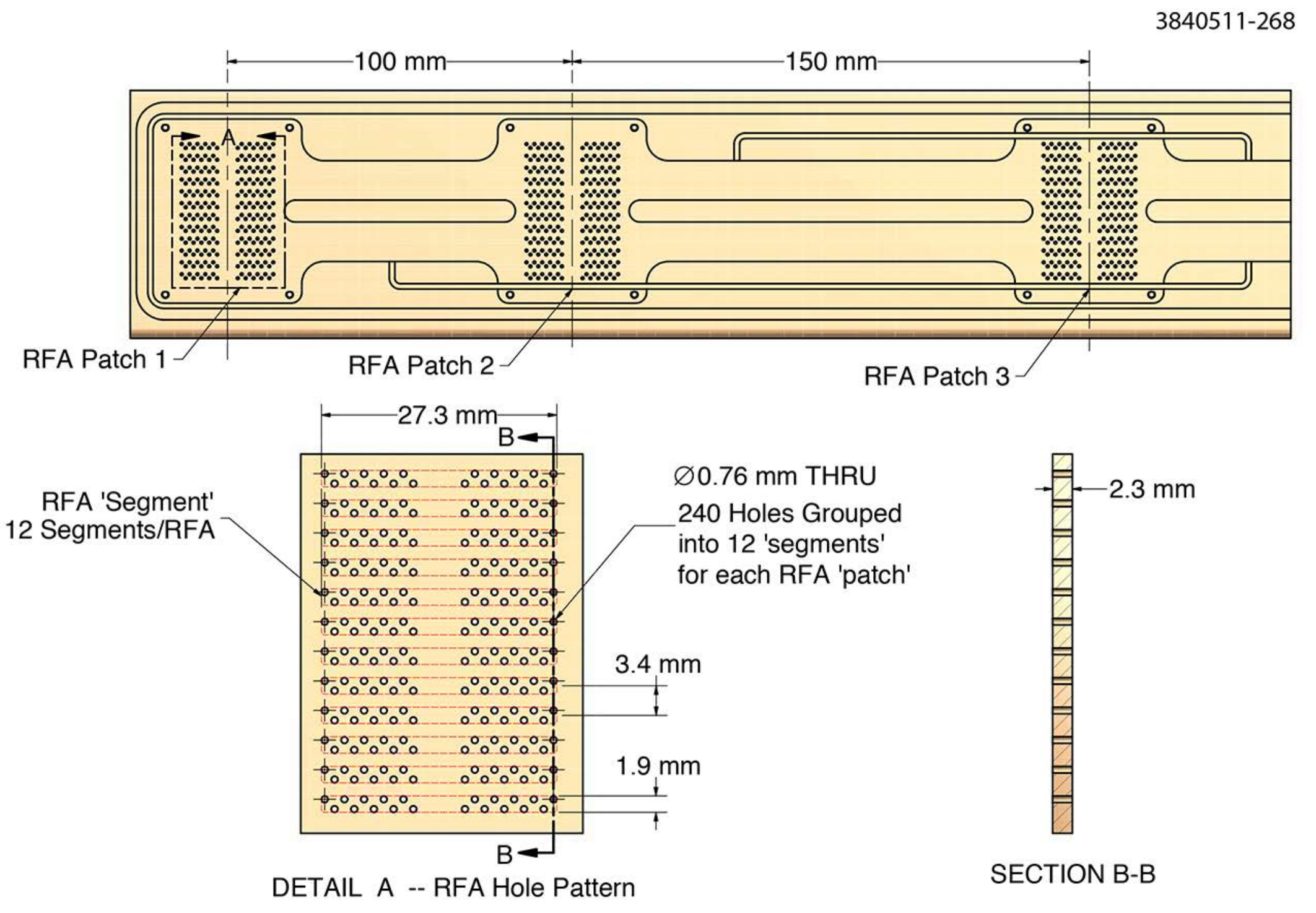}
	\caption[RFA hole patterns for the SCW RFA beam pipes.] {Small holes are drilled through top beam pipe to allow electrons in the beam pipe pass into RFAs. There are 240 holes for each RFA, and they are grouped into 12 segments to sample transverse EC density distribution.  A backing plate was used in drilling these holes to avoid burrs.\label{fig:SCW_RFA_Holes}}
\end{figure}

	Electrons from the EC in the beam pipe pass through transmission holes to the RFA detector.  There are 240 small holes in each of the three RFAs.  As shown in Figure~\ref{fig:SCW_RFA_Holes}, they are grouped into 12 segments in the transverse direction to the beam pipe.  These holes have a diameter of 0.75~mm ($\frac{1}{3}$ of the wall thickness) to reduce noise from the beam's radio frequency (RF) electromagnetic interference (EMI.)

	The retarding grids are nested in ceramic frames with the frames bolted to the RFA housing pockets.  Two generations of metallic meshes were used as the retarding grids.  The first generation was made of photo-chemically etched 0.15~mm-thick stainless steel (SST) mesh, with an optical transparency of approximately 38\%.  
The SST mesh is easy to handle and inexpensive.  However, the SST mesh has two drawbacks.  Its relatively low transmission not only directly translates to low EC detection efficiency, but also limits the EC diagnostic accuracy due to significant interaction of the transmitted electrons with the grid.  To reduce the secondary emission from the grid, the SST meshes were coated with approximately 0.3~$\mu$m of gold.  The second generation grids (by Precision EForming, Inc) were electro-formed copper meshes, bonded to SST frames (supplied by Cornell).  The electro-formed copper meshes were also coated with gold (approximately 0.3~$\mu$m in thickness) via electroplating to reduce secondary electron emission.  The electro-formed meshes consist of 15~$\mu$m wide and13~$\mu$m thick copper wires with spacing 0.34~mm in both transverse directions and an optical transparency of approximately 92\%.  
The disadvantage of the copper electro-formed mesh is that it is very expensive and fragile.

\begin{figure}
	\centering
	\includegraphics[width=0.75\textwidth]{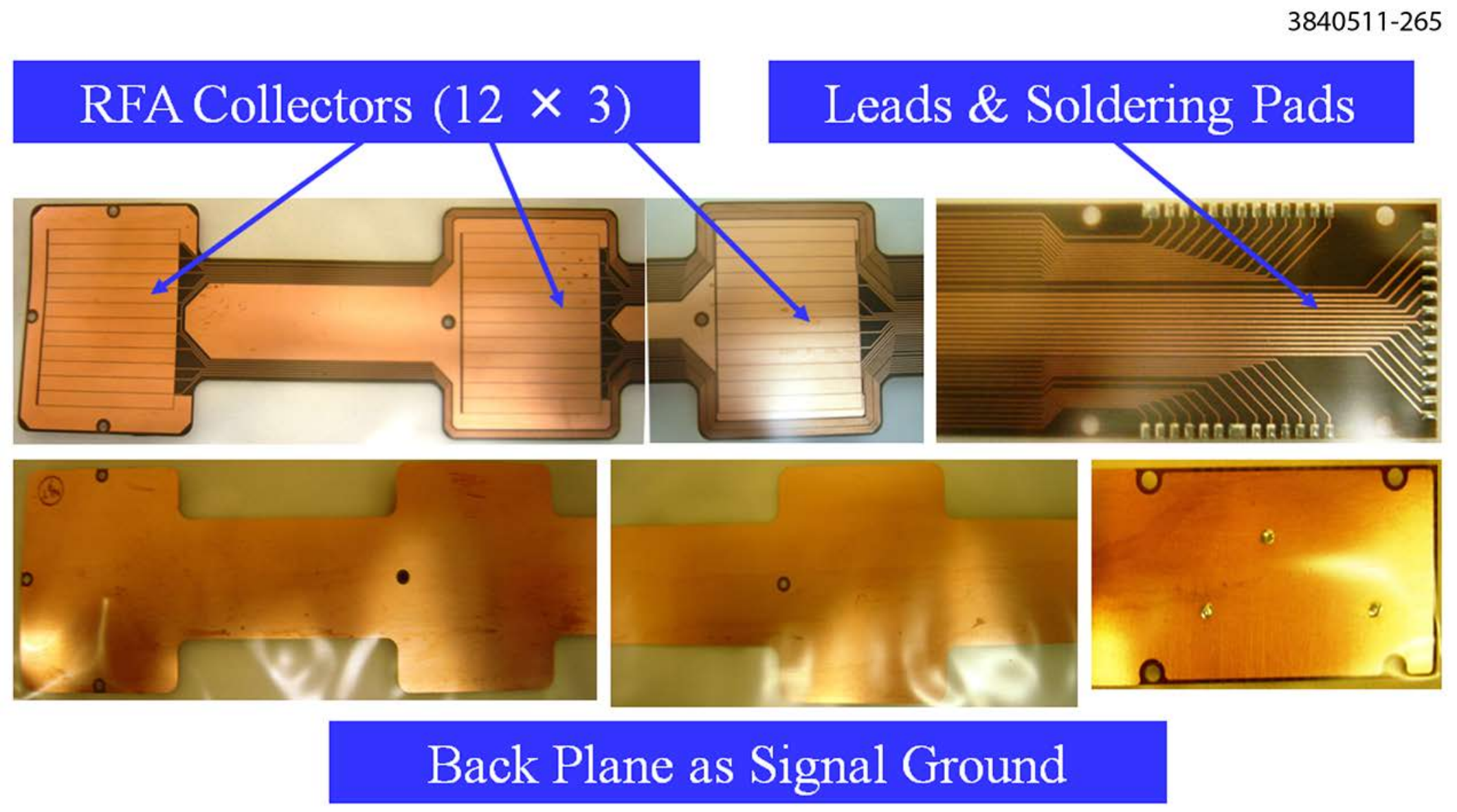}
	\caption[Flexible circuit used as wiggler RFA collectors]{Photos of the flexible circuit used as RFA electron collectors in the SCW RFA beam pipe.  The flexible circuit is made of a thin Kapton sheet with copper cladding on both sides, having a thickness of 0.15 to 0.20~mm and length of 886~mm. \label{fig:SCW_RFA_circuit}}
\end{figure}
\begin{figure}
	\centering
	\includegraphics[angle=270,width=0.5\textwidth]{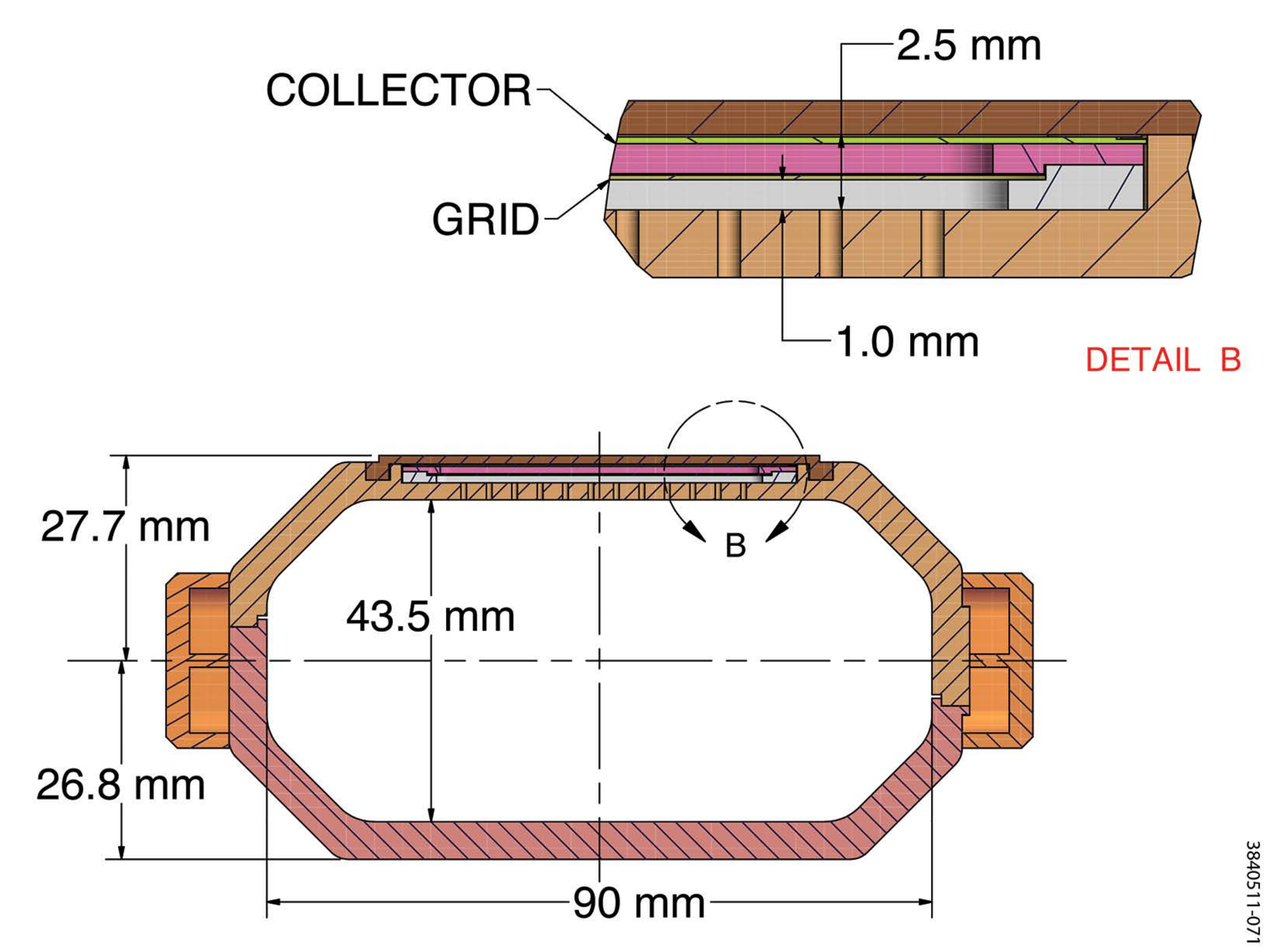}
	\caption{Cross section view of a RFA structure on the SCW beam pipe \label{fig:SCW_RFA_cross_section}}
\end{figure}

	Similar to the thin RFA design\cite{NIMA760:86to97}, a flexible copper-clad/Kapton circuit was used as RFA electron collector.  As shown in Figure~\ref{fig:SCW_RFA_circuit}, three sets of RFA pads each contain 12 transverse collectors to match the hole patterns on the RFA beam pipe. After tinning the soldering pads with Sn(63\%)Pb(37\%) solder, the flexible circuits were cleaned and degassed with a 150$^\circ$C vacuum bakeout before insertion in the RFA beam pipe.  Vacuum measurements (with a residual gas analyzer) showed acceptable vacuum properties for the clean flexible circuits. The flexible circuit strip was laid on top of the ceramic frames of the grids and precisely positioned with ceramic head-pins.  After positioning the flexible circuit and feeding its connection pad through a `duck-under' channel (see `Detail B' in Figure~\ref{fig:SCW_RFA_structure}) into the RFA connection port, ultra high vacuum- (UHV-) compatible Kapton-coated copper wires (42 wires per assembly) were soldered to the connection pads.  These wires were also pre-tinned and cleaned so that the entire RFA circuit is free from solder flux.  The final RFA connections are wired to three 15-pin D-type vacuum feedthroughs on a 4.5~inch ConFlat$^\copyright$ flange.  A copper RFA vacuum cover was electron-beam welded to the beam pipe's top half to complete the RFA assembly.  Figure~\ref{fig:SCW_RFA_cross_section} shows the cross-section of the RFA beam pipe at one of the RFA locations.  The entire RFA structure is contained within a total vertical space of 2.5~mm.

\subsection{Preparation of SCW for RFA Beam Pipe}

\begin{figure}
	\centering
	\includegraphics[width=0.8\textwidth, angle=-90, width=6.0in]{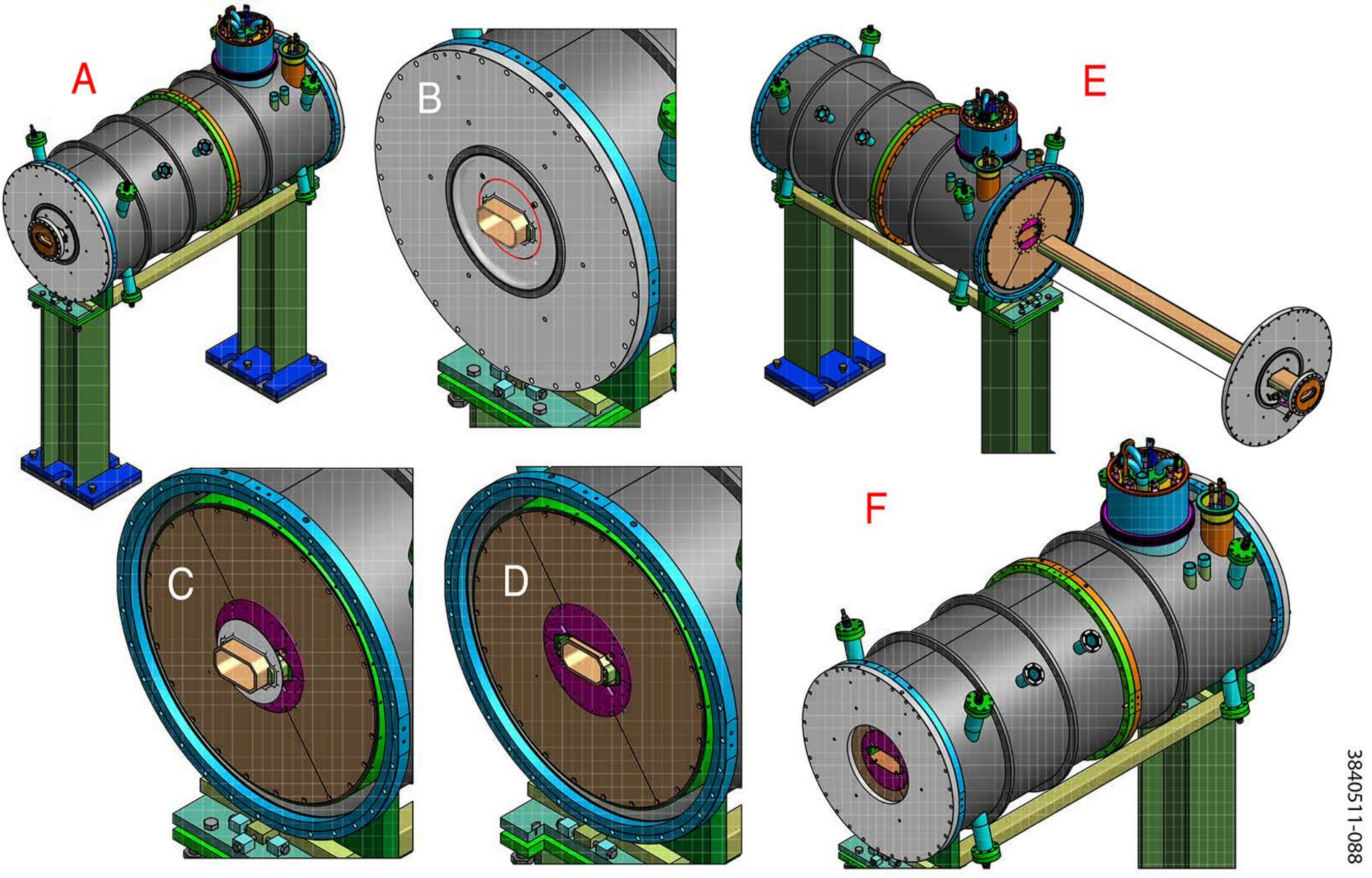}
	\caption[Sequence of the extraction of the beam pipe from CESR-c SCW Assembly.]{Sequence of the extraction of the beam pipe from CESR-c SCW Assembly. (A) One of end beam pipe flanges is cut off with a saw; (B) A circular cut is made through the flexible disk (indicated by red line) with a hole saw; (C) The large insulation vacuum vessel flange is demounted; (D) Another saw cut is made on the beam pipe behind the remainder of the flexible disk; (E) The remaining beam pipe is then pulled out of the SCW's warm-bore using the large flange at other end; (F) Both large vacuum vessel end flanges are remounted, after some surface machining to remove remnants of the flexible disks.
\label{fig:SCW_beampipe_extraction}}
\end{figure}

Preparations must be undertaken for a CESR-c SCW to receive an RFA beam pipe assembly.  The main task is to extract the existing beam pipe from a CESR-c SCW without disturbing the magnet's structure.  As shown in Figure~\ref{fig:SCW_CESR_c}, the CESR-c SCW beam pipe is attached to the insulating vacuum vessel via two thin stainless steel disks.  The major portion of these disks is about 1.5~mm thick.  Called flexible disks, they allow for differential thermal expansion and contraction of the beam pipe relative to the vacuum vessel.  The flexible disks completed the insulation vacuum envelope after being welded to the beam pipes and the large end flanges of the vacuum vessel.  The sequence of the beam pipe extraction from a CESR-c SCW is described in Figure~\ref{fig:SCW_beampipe_extraction}.  After the extraction of the beam pipe, all debris from the extraction operation was thoroughly removed from the magnet and a pair of large end flanges (with O-ring seals) were mounted to the insulation vacuum vessel. Then the vacuum vessel was leak checked.  The integrity of the wiggler magnet was also verified. 

\subsection{RFA Beampipe Fabrication and Integration to SCW}

This section describes the hardware construction and installation for and RFA in a bare copper beam pipe.
Many steps of heavy welding are required as part of the RFA wiggler beam pipe construction.  As illustrated 
in Figure~\ref{fig:SCW_RFA_structure}, we have designed the equipment in a such way so as 
to have all the welding that may overheat the portions of the beam pipe near the RFA flexible circuit 
completed prior to the installation of the flexible circuit.  This Kapton$^\copyright$-based circuit has a 
temperature rating of 220$^\circ$C.  To address this requirement, a `duck-under' channel was created beneath 
the stainless steel flexible disk, which later was welded to the wiggler insulation vacuum vessel.  During 
the construction process all vacuum welds, except the final RFA vacuum cover, can be completed and leak checked 
prior to the installation of the heat-sensitive flexible circuit.  Utilizing the `duck-under' channel, one can 
feed the flexible circuit from the RFA portion of the beam pipe (i.e. inside wiggler insulation vacuum space) to 
the RFA connection port.

\begin{figure}
	\centering
	\includegraphics[width=0.75\textwidth]{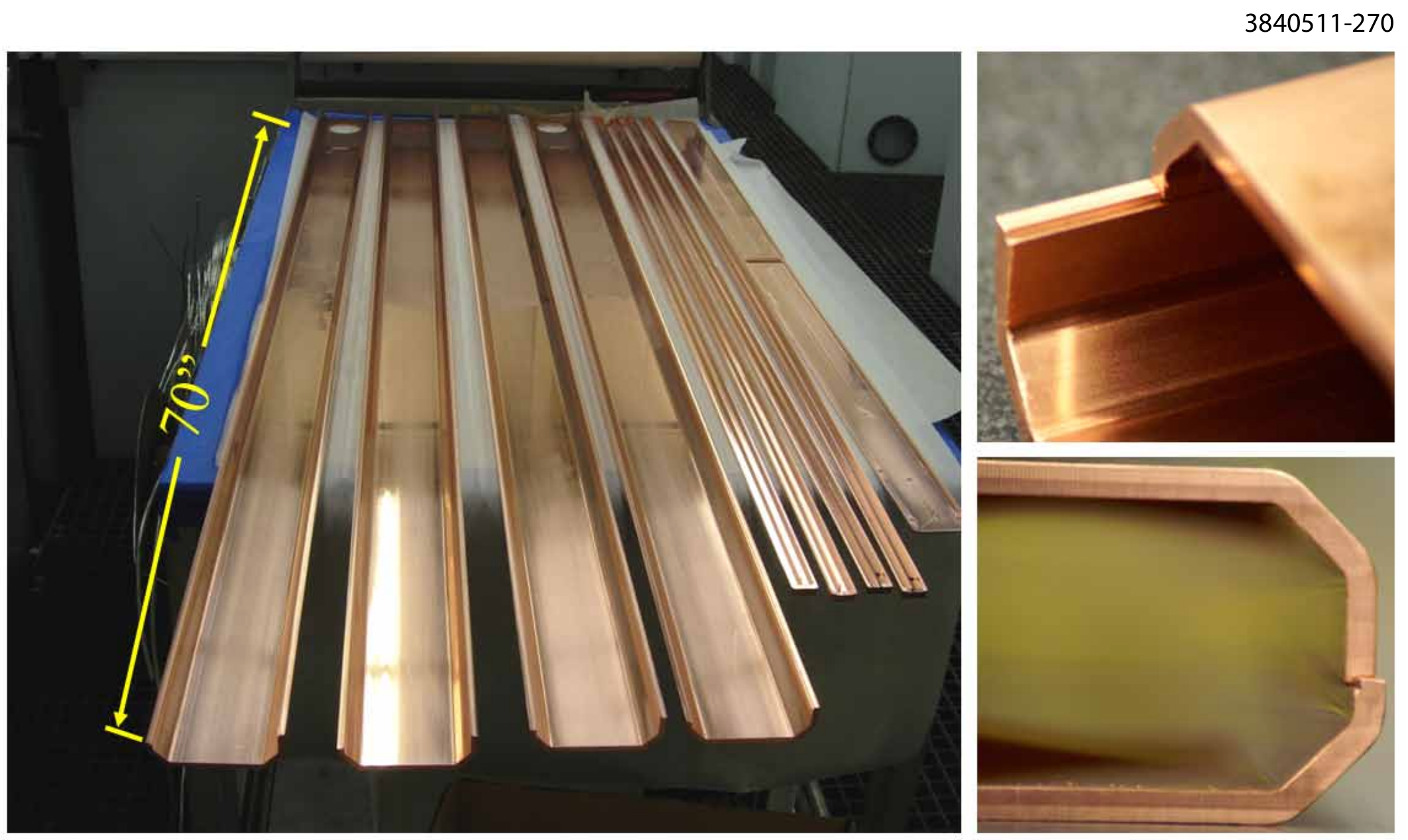}
	\caption[Wiggler beam pipe Machined photos]{The copper beam pipes are split into two halves with EB-weld preps ready for cleaning and EB-welding.  (Photos: courtesy of Dawn Munson of LBNL) \label{fig:SCW_RFA_pipe_machined}}
\end{figure}

\begin{figure}
	\centering
	\includegraphics[width=0.65\textwidth]{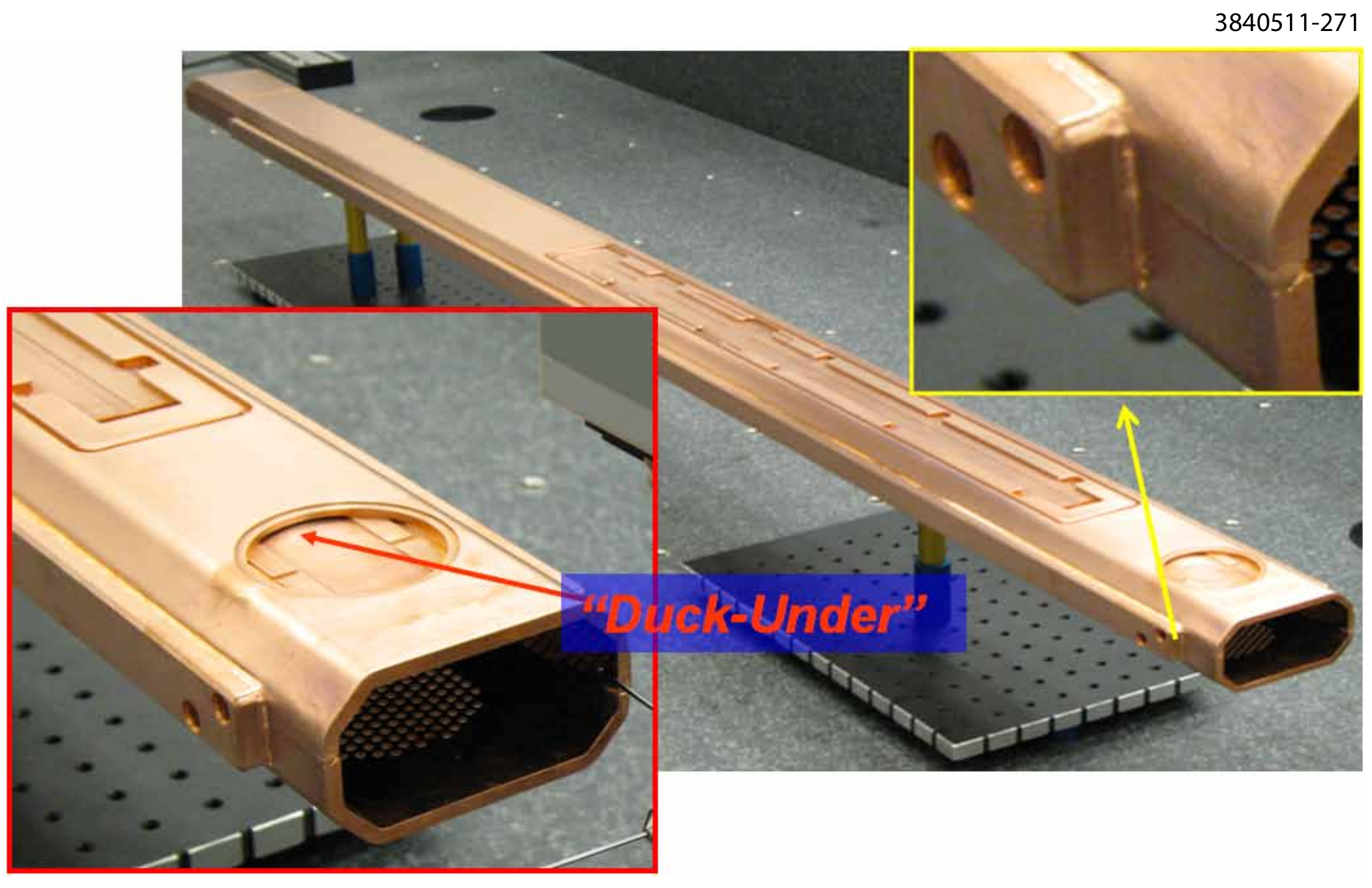}
	\caption[Wiggler beam pipe EB-welded Photos]{The beam pipe halves were EB-welded with approximately a 1~mm beam
penetration.  Two side cooling channels were also EB-welded to the beam pipes.  The inserts show close-up views of the `duck-under' channel for the RFA flexible circuit (lower right) and the EB-welds (upper left).  The welded beam pipe is shown here on a coordinates measuring machine (CMM), which gave the total measured distortion of less than 0.15~mm.  (Photos: courtesy of Dawn Munson of LBNL) \label{fig:SCW_RFA_pipe_welded}}
\end{figure}

The beam pipe fabrication itself began by splitting the fully annealed oxygen free electronic (OFE) copper extrusion into two halves.  The copper extrusions are the same type used for PEP II Low Energy Ring (LER) Quad beam pipe and the original CESR-c SCW beam pipes.  All RFA-related features on the top half were machined at LBNL's machine shop with computer numerically controlled (CNC) machinery.  These include the RFA grid pockets, 720 electron transmission holes, the `duck-under' channel, and all the EB weld-preps (see photos in Figure~\ref{fig:SCW_RFA_pipe_machined}).  After cleaning, an `under cover plate' was EB welded to the top half of beam pipe to form the `duck-under' channel.  Then two finished halves were joined together using a CNC EB welder, with approximately a 1~mm electron-beam penetration at the seams.  After passing a leak check of the beam pipe seams, two side cooling channels were then EB welded to the beam pipe (see photos in Figure~\ref{fig:SCW_RFA_pipe_welded}).  Measurements using numerically controlled (NC) coordinate machine found that amount of distortion of the EB-welded beam pipe is well within design tolerances.  Subsequently, all other vacuum components (including end flanges and transitions, RFA connection port and vacuum pumping port) were manually welded to the beam pipes with tungsten inert gas (TIG) welding in an argon atmosphere to avoid the oxidation of the copper beam pipes.  A temporary flange was welded to the end of the beam pipe, which is away from the RFA locations, to facilitate initial leak checking, bakeout and TiN coating.  At this stage all the UHV joints of the beam pipes, with exception of the top RFA vacuum cover plate, were finished and the beampipe was ready for the final RFA component assembly. To ensure the required UHV quality, both beam pipes were baked to 150$^\circ$C under vacuum. After the bakeout, one of the two RFA beam pipes produced at LBNL was coated with TiN thin film on the beam pipe's interior by the SLAC team.  By comparing with the TiN coated and bare copper chambers using the {\cesrta} diagnostics, the effectiveness of TiN coating in suppressing EC in wiggler magnetic fields could be evaluated.  

\begin{figure}
	\centering
	\includegraphics[width=0.8\textwidth]{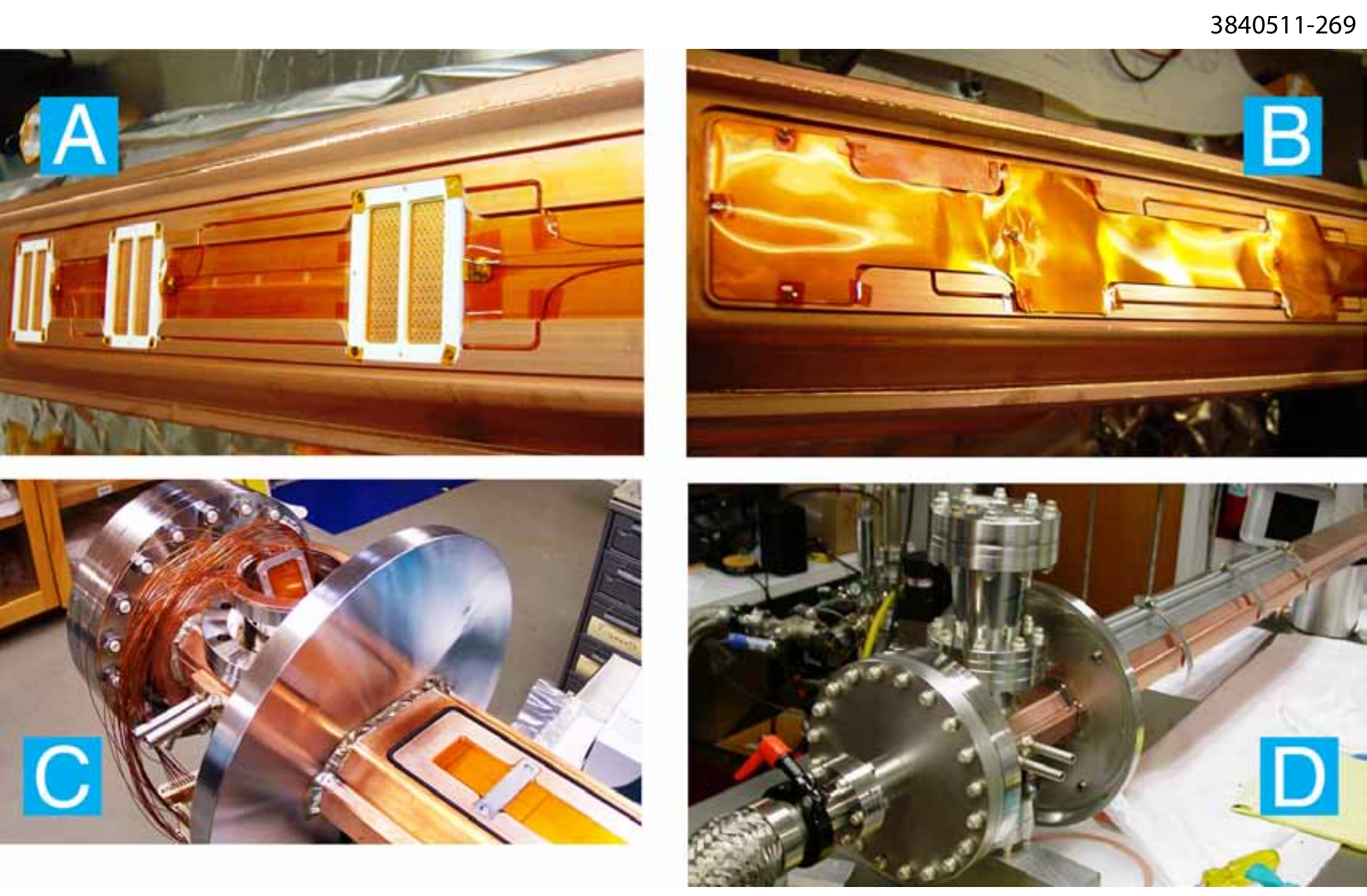}
	\caption[Photos showing RFA installation on a wiggler beam pipe]{Photographs of the key steps in the RFA installation on a wiggler beam pipe: (A) Three grids are installed and individually wired to the connection port; (B ) The flexible circuit collector is installed and located with 5 ceramic head-pins; (C) With the circuit through the `duck-under' tunnel, all signal wires are attached in the connector port; (D) After making the final RFA connections, a vacuum leak-check is performed and a final RFA electrical check-out is conducted with the chamber under vacuum before EB-welding the RFA cover. \label{fig:SCW_RFA_Photos}}
\end{figure}

The first two partially finished wiggler beam pipes at LBNL were shipped to Cornell (in September 2008) for final RFA component assembly and integration into the SC wiggler assemblies.  Photos in Figure~\ref{fig:SCW_RFA_Photos} show some of the key steps of the RFA installation.  The entire RFA installation was performed in a Class 1000 Clean Room.  After thorough electrical checks the RFA installation was finished by E-beam welding on the OFE copper RFA vacuum cover.  The finished RFA beam pipes were baked to 150$^\circ$C for 48~hours before final insertion into the prepared wiggler magnet assemblies.  

Before integrating the RFA beam pipe into the SCW, the temporary flanged end (end away from the RFAs) was machined off using a clean milling machine with clean cutting tools.  Extreme care was taken in this final machining step, which included blocking metallic debris from entering the RFA section of the beam pipe and constant purging of the chamber with N$_2$.  The RFA beam pipe was inserted into the wiggler warm-bore and precisely positioned with respect to the wiggler magnet by optical survey.  Extreme care was also taken in the final welding stages (including final beam pipe flange and seals to the wiggler insulation vacuum vessel) to prevent overheating of the RFA components.

\begin{figure}
	\centering
	\includegraphics[width=0.75\textwidth]{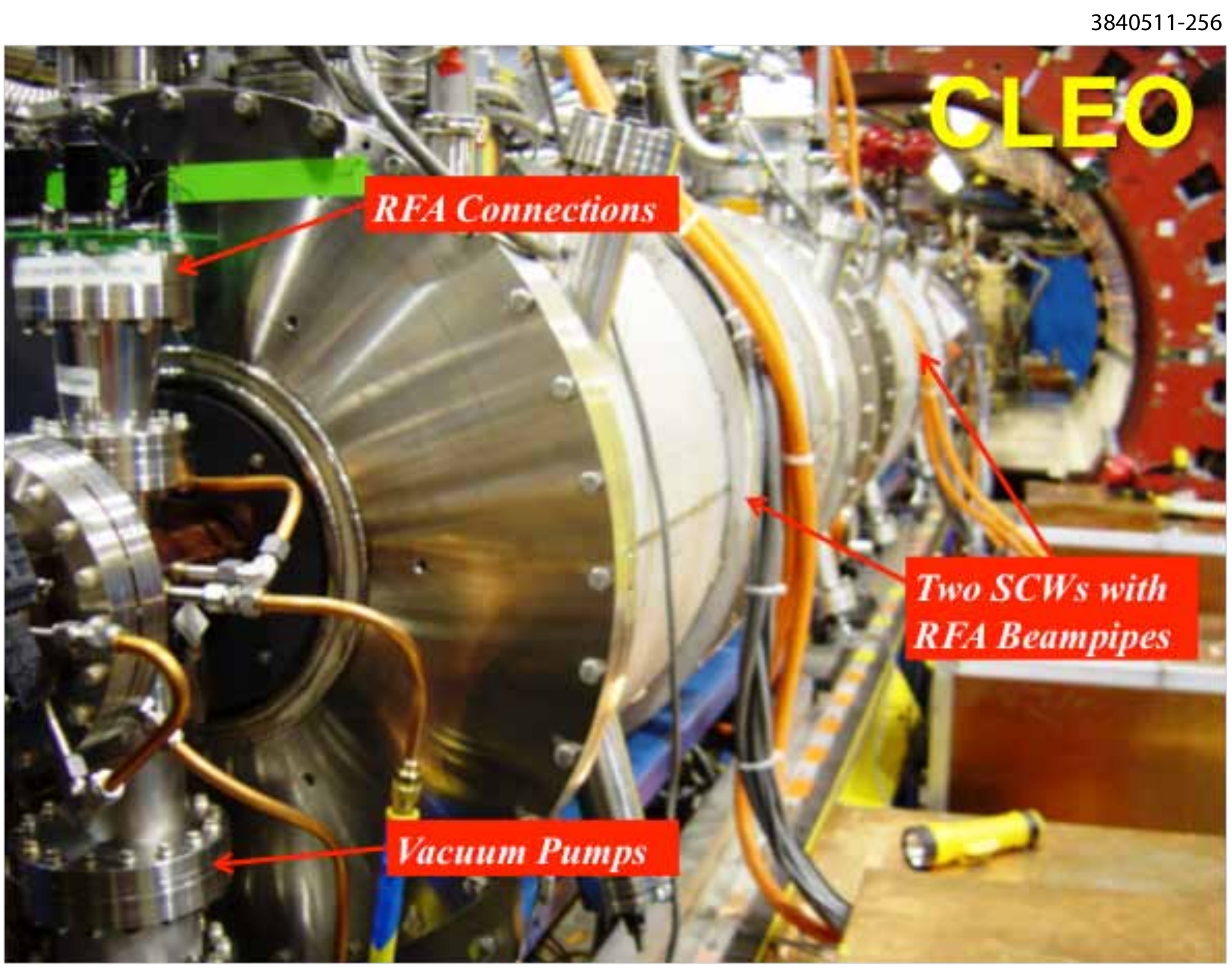}
	\caption{SCWs with RFA beam pipes were installed in the L0 Experimental region. \label{fig:SCW_RFA_in_L0}}
\end{figure}

\begin{figure}[hbt]
    \centering
    \includegraphics[width=\textwidth]{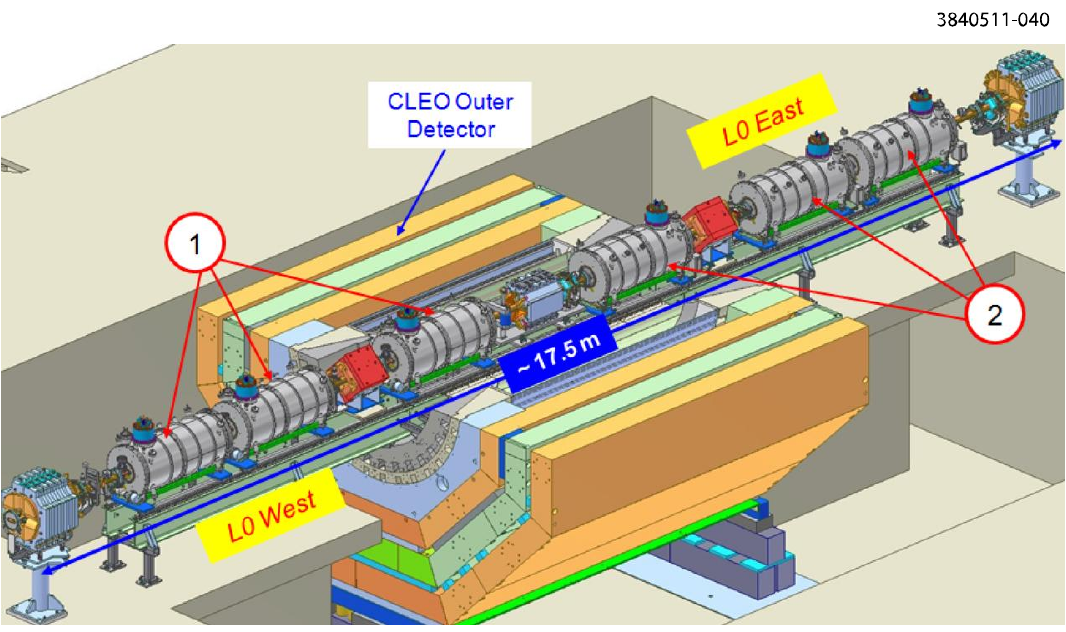}
    \caption[L0 EC Experimental Region Vacuum Layout]{L0 center {\cesrta} EC experimental region, consists of (1) three RFA-equipped SCWs and (2) three CESR-c SCWs.  Many other EC diagnostics, such as RFA in the drifts, BPMs and TE-Wave buttons, are also implemented.  The six SCWs in the region are labelled, from west to east, SCW02WB, SCW02WA, SCW01W, SCW01E, SCW02EA and SCW02EB\cite{JINST10:P07013}.\label{fig:vac_l0_center}}
\end{figure}

Finally, the completed SCWs with RFA beam pipes were `baked' at 70$^\circ$C for 2 days by circulating hot water through the beam pipe cooling channels in an effort to de-gas the vacuum components after the prolonged air exposure during the final RFA beam pipe insertion and welding.  The first two SCWs with RFA beam pipe were successfully installed in the west side of the L0 Experimental region in CESR in October 2008 (see Figure~\ref{fig:SCW_RFA_in_L0}) in the SCW02WA and SCW02WB locations (refer to Figure~\ref{fig:vac_l0_center}).

\subsection{Grooved Chamber}

\begin{figure}
	\centering
	\includegraphics[width=0.8\textwidth, angle=-90, width=6.0in]{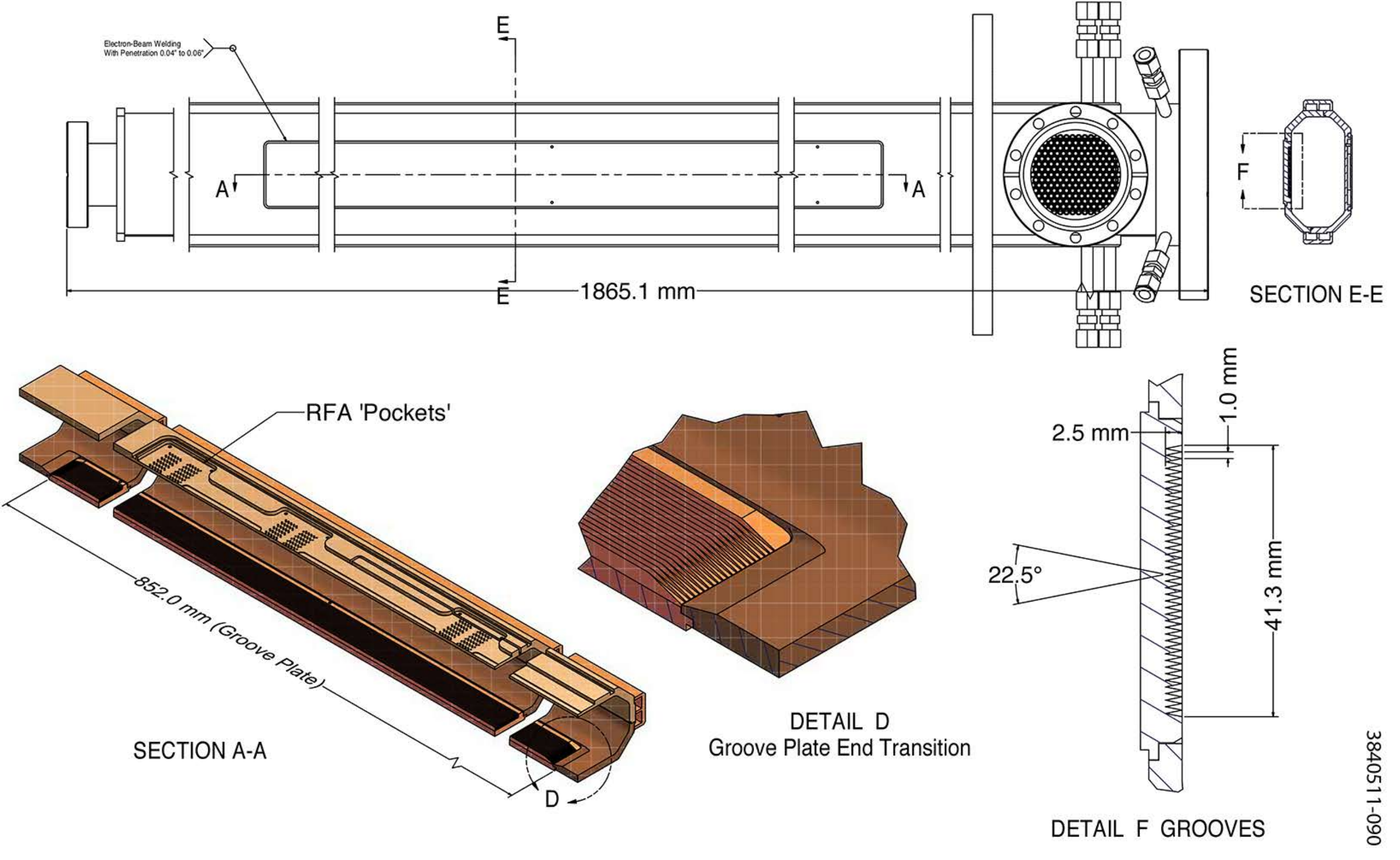}
	\caption{SCW RFA beam pipe with a groove plate EB welded to the bottom beam pipe. \label{fig:SCW_RFA_pipe_grooves}}
\end{figure}

Using grooved surfaces to lower effective secondary emission yield (SEY) is a well-known \cite{ECLOUD04:139to141, JAP104:104904, VACUUM73:195to199, NIMA571:588to598} passive technique to suppress EC growth in a magnet field.  Having successfully implemented RFAs into SCWs in CESR, we studied the effects of grooved chambers by constructing an RFA-equipped SCW chamber with a grooved insert installed on the bottom surface.  Figure~\ref{fig:SCW_RFA_pipe_grooves} shows the design of the RFA beam pipe.  This beam pipe assembly is basically identical to the design as shown in Figure~\ref{fig:SCW_RFA_structure} with the exception that a copper plate with triangular grooves has been attached to the bottom half via EB-welding.  The same procedure (described above) was followed for the RFA installation to the beam pipe and the beam pipe integration into the SCW.

The direct implementation of a grooved surface in a copper extrusion was found to be quite challenging.   Practical considerations dictated the implementation of the grooved surface using an inserted plate.  First, it was determined (through test machining) that the copper extrusions used for the SCW beam pipe were fully annealed, and too soft for machining the sharp-tipped grooves.  Thus the grooves were machined in a separate plate of full-hard OFHC copper which could be electron-beam welded into the vacuum chamber instead of directly being machine onto the bottom half extrusion.  However, even in the case of machining a separate plate, it was too costly to machine grooves for the entire length of the chamber.  Thus, for the experimental tests a grooved plate of sufficient length to span the RFAs was used.

\begin{figure}
	\centering
	\includegraphics[width=0.75\textwidth]{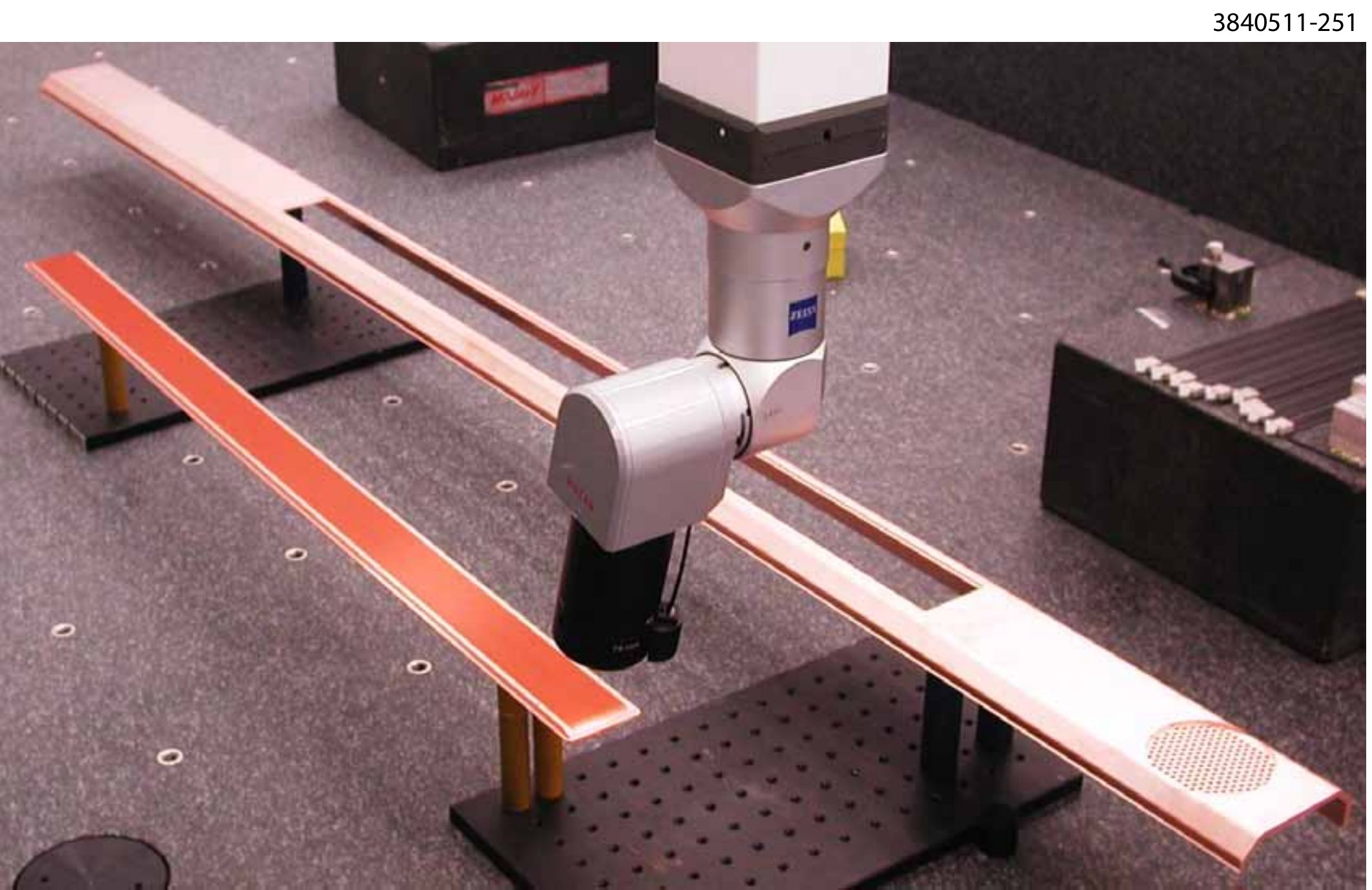}
	\caption{Bottom half of an SCW beam pipe with a cut-out and the groove plate during inspection on a CMM (Photo courtesy of Dawn Munson of LBNL) \label{fig:SCW_btm_groove}}
\end{figure}

The geometry and dimensions of the grooves are shown in the `Detail F' in Figure~\ref{fig:SCW_RFA_pipe_grooves}.  A cut-out opening was machined for the grooved plate into the bottom half of the beam pipe (Figure~\ref{fig:SCW_btm_groove}).  Sloped cuts of 30$^\circ$ were made on both ends of the groove plate and on the beam pipe wall (shown in `Detail D' in Figure~\ref{fig:SCW_RFA_pipe_grooves}), which provide a smooth transition from the flat surface to the triangular groove tips in order to minimize higher order mode loss (HOML.)

\begin{figure}
	\centering
	\includegraphics[width=0.5\textwidth]{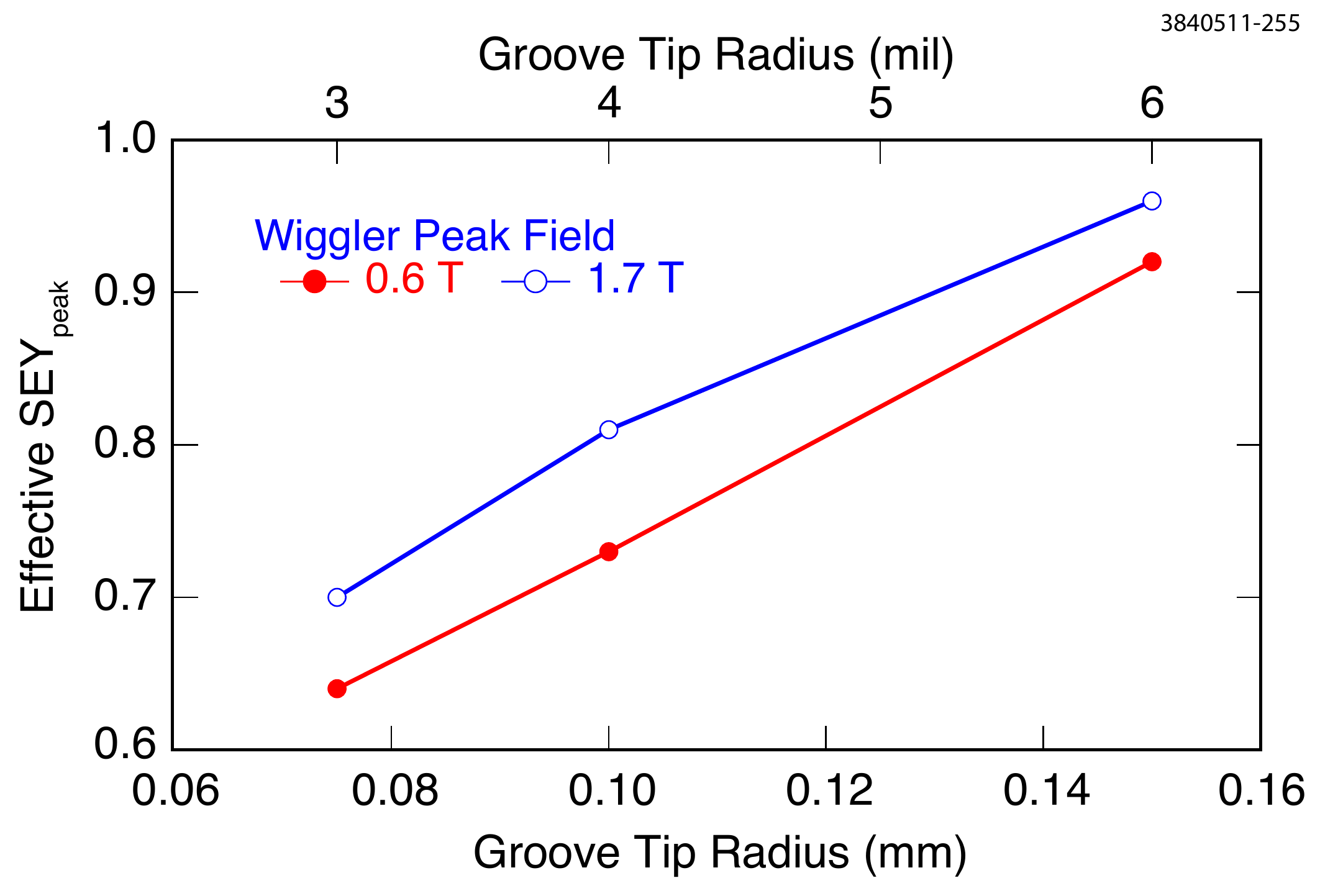}
	\caption[Simulated effective SEY on triangular grooves in SCW]{Simulated effective peak SEY on triangular grooves on copper as a function of groove tip/valley radius (sharpness) for two typical wiggler peak fields.  (The simulation assumed the tips and valleys to have the same radius.)  The two peak wiggler fields correspond to CESR beam energies of 1.8~GeV and 5.2~GeV.\label{fig:SCW_groove_tip_simulation}}
\end{figure}

\begin{figure}
	\centering
\begin{tabular}{cc}
\includegraphics[width=0.6\textwidth]{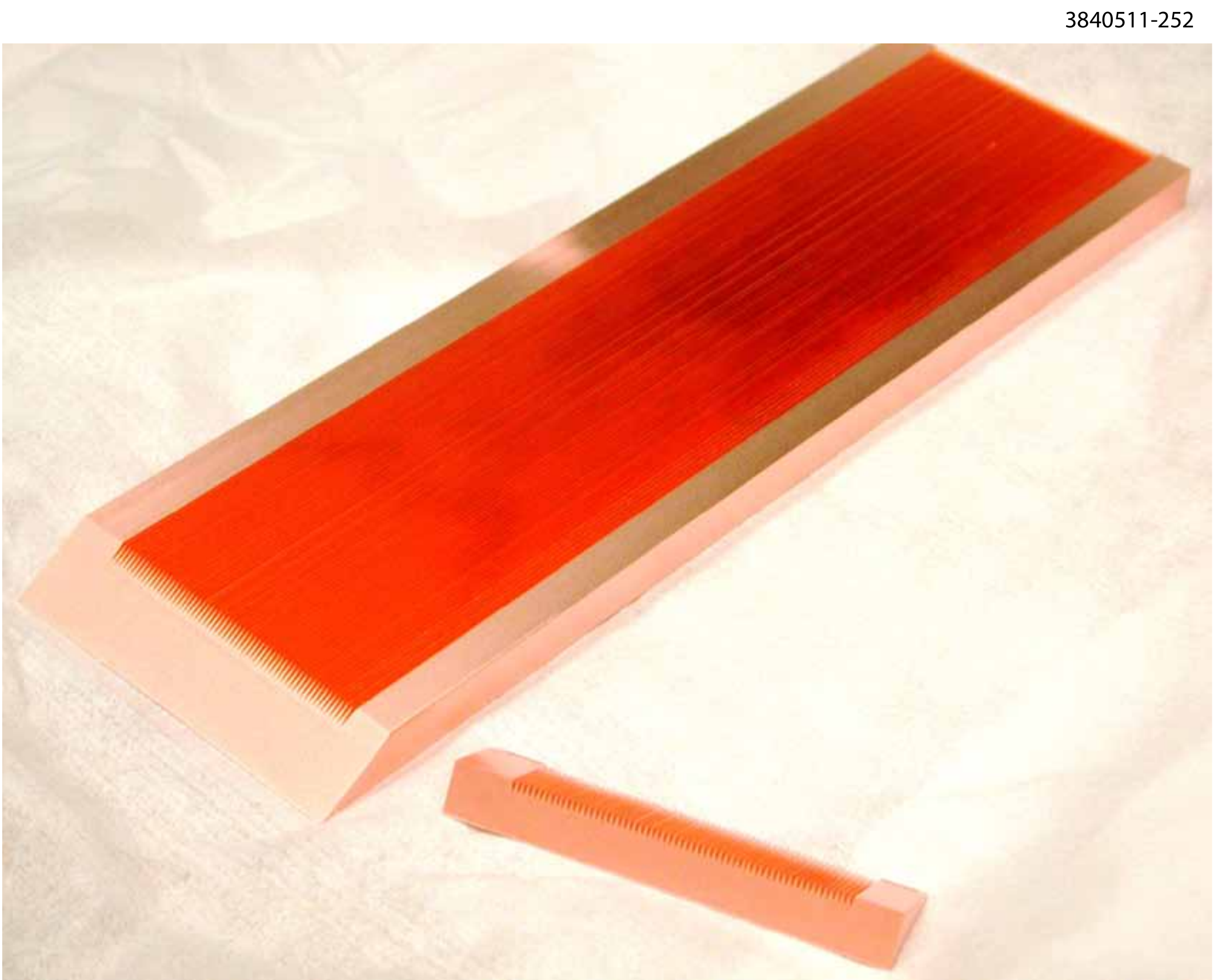} &
\includegraphics[width=0.4\textwidth]{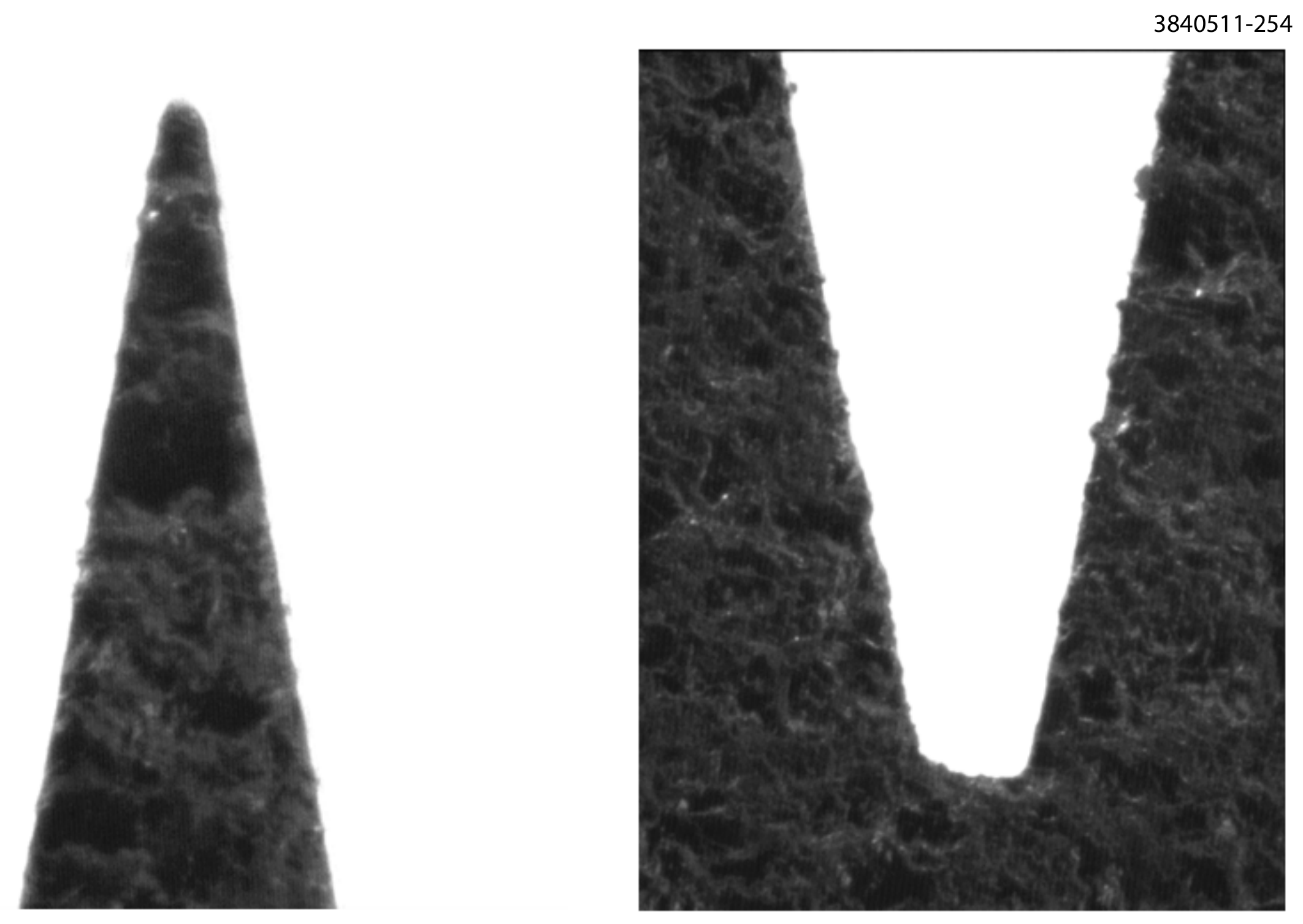}\\
\end{tabular}
	\caption[Photo of a test SCW groove plate and Tip/Valley Inspection]{LEFT: A prototype groove plate was fabricated with full-hard OFC copper.  A precision wire-cut across the grooves was performed for groove sharpness inspection.  RIGHT: Images of prototype groove tip and valley, taken with a high-resolution camera on the CMM at LBNL with back-lighting. The tip radius is estimated to be approximately 0.001" (25~$\mu$m) and the valley approximately 0.0025" (63~$\mu$m).  (Photo courtesy of Dawn Munson of LBNL)  
\label{fig:SCW_groove_test}}
\end{figure}

The triangular grooves were made with a milling technique using specially designed cutters.  In practice there will be finite radius on the tips and valleys of these grooves.  The effective peak SEY on these triangular grooves was simulated as a function of the tip/valley radius by Wang \cite{LWang:PrivateC}.  As shown in Figure~\ref{fig:SCW_groove_tip_simulation}  for the grooved surfaces to be effective in suppress SEY, it is essential to produce these triangular grooves with average radius of their tips and valleys smaller than 0.003" (75~$\mu$m).  To qualify this groove fabrication technique, a prototype groove plate (of same width, but 150~mm long) was produced for inspection.  Figure~\ref{fig:SCW_groove_test} shows the prototype groove plate, and images from optical inspections.  The resulted tip and valley radii were approximately 25~$\mu$m and 60~$\mu$m, respectively, which were satisfactory.

The RFA beam pipe assembly, having the RFA features on the top half and the grooved plate on the bottom half, was fabricated at LBNL and delivered to Cornell in May 2009.  The thin RFAs were assembled into the beam pipe and then the RFA beam pipe was inserted into a SCW, following a similar procedure as describe previously in this section.  Extra measures were taken to protect the sharp grooves at the bottom of the beam pipe throughout the final stages of assembly.  The completed SCW was successfully installed for the {\cesrta} program \cite{JINST10:P07012, JINST10:P07013, JINST11:P04025} in the L0 EC experimental region in July 2009 in the SCW02WB location (refer to Figure~\ref{fig:vac_l0_center}).

The SCW beam pipe assembly, containing the grooves and their associated RFAs, was successfully operated through two {\cesrta} experimental runs (July~2009 to September~2009, and November~2009 to December~2009) and two CHESS runs (September~2009 to November~2009, and January~2010 to March~2010), accumulating total beam doses of approximately 940~Amp$\cdot$hr.  During the April~2010 CESR shutdown, this SCW assembly was replaced with a SCW assembly fitted with RFA beam pipe, having an EC clearing electrode (see the subsequent sub-section).  After its removal from CESR the following procedures were performed on the grooved RFA beam pipe as a continuing effort in studying EC suppression in wigglers.

\begin{figure}
	\centering
	\includegraphics[width=0.75\textwidth]{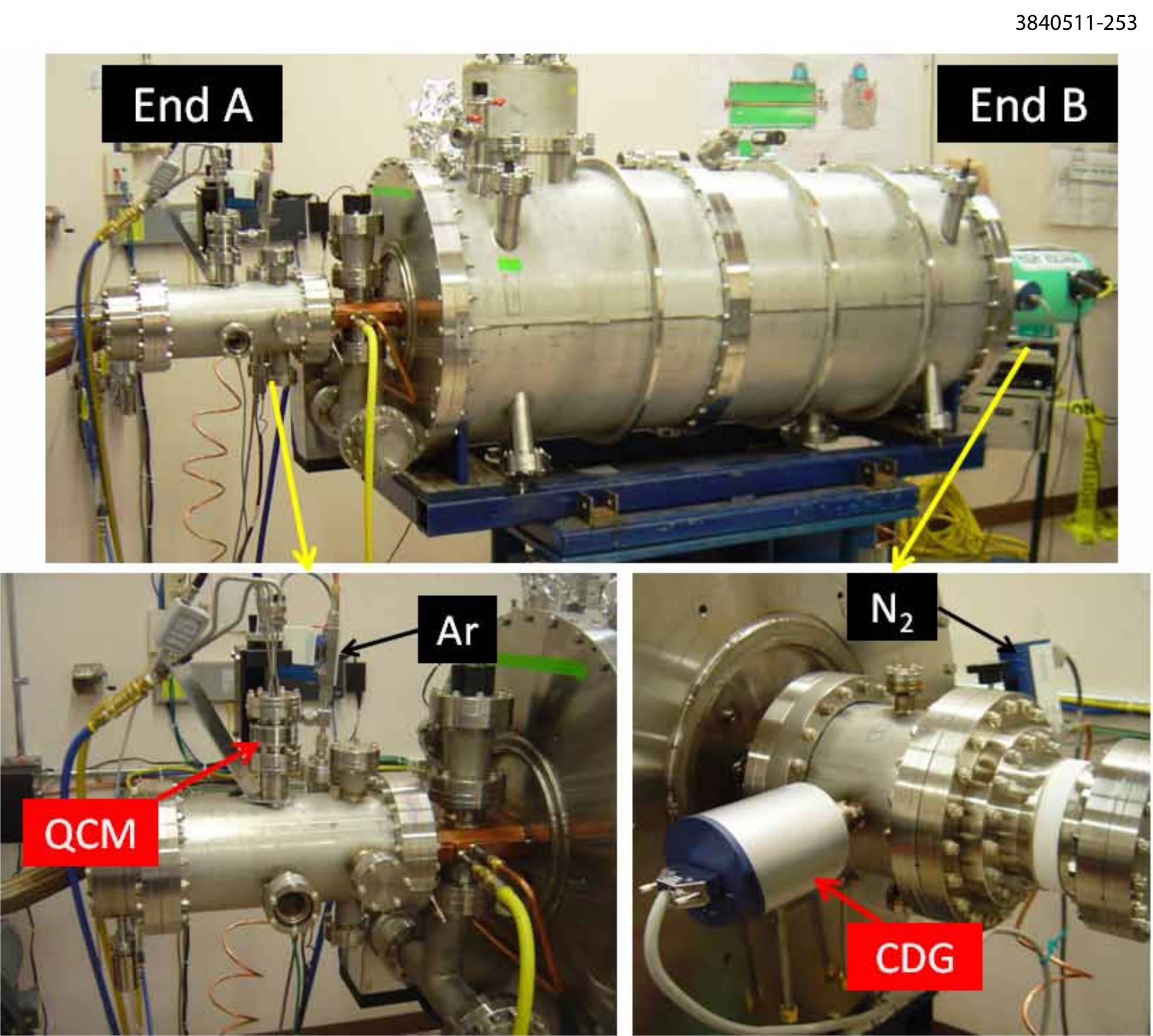}
	\caption[TiN coating setup for SCW RFA beam pipe with grooved bottom plate]{TiN coating setup of the SCW RFA beam pipe with bottom grooved plate.  Solid grade 1 titanium rods were used as cathode, biased at a DC voltage of 1.5 kV.  Flow rates of sputtering gases (Ar and N$_2$) were controlled with flow meters, and the sputtering pressure was controlled at approximately 0.10~$\rm{torr}$.  A Quartz Crystal Micro-balance (QCM) was used to monitor film thickness. \label{fig:SCW_groove_TiN_setup}}
\end{figure}

\begin{figure}
	\centering
	\includegraphics[width=0.5\textwidth]{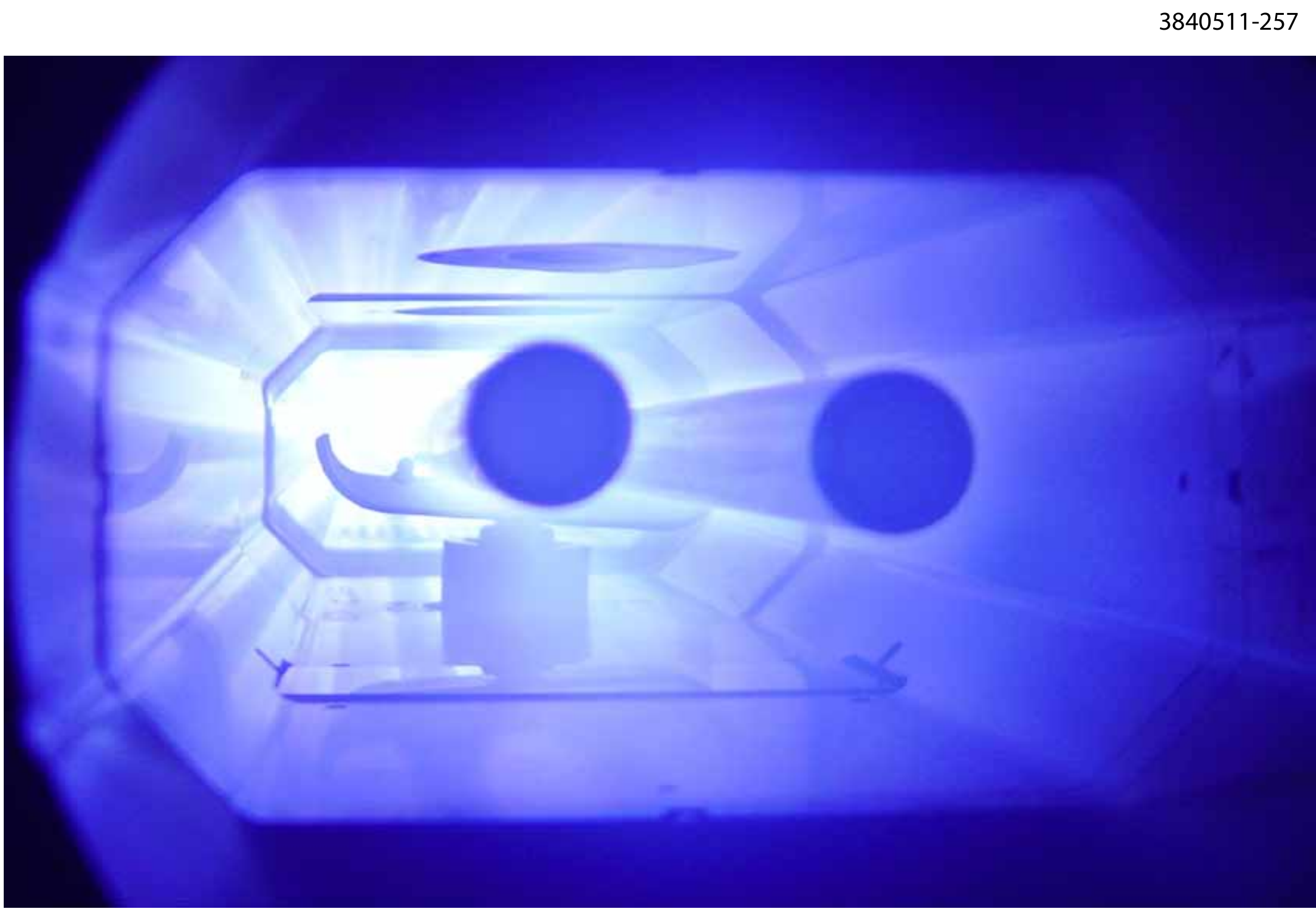}
	\caption[Glow discharge pattern during TiN coating of SCW RFA beam pipe with bottom grooved plate.]{Glowing discharge during TiN coating of the grooved RFA beam pipe. The coating was performed in two stages, the first stage using two Ti rods (as shown) and the second stage with three rod (located in different transverse positions.)  The two-stage coating is to minimize shadowing effects from the steeply grooved troughs. \label{fig:SCW_TiN_glow}}
\end{figure}

\begin{figure}
	\centering
\begin{tabular}{cc}
\includegraphics[width=0.45\textwidth]{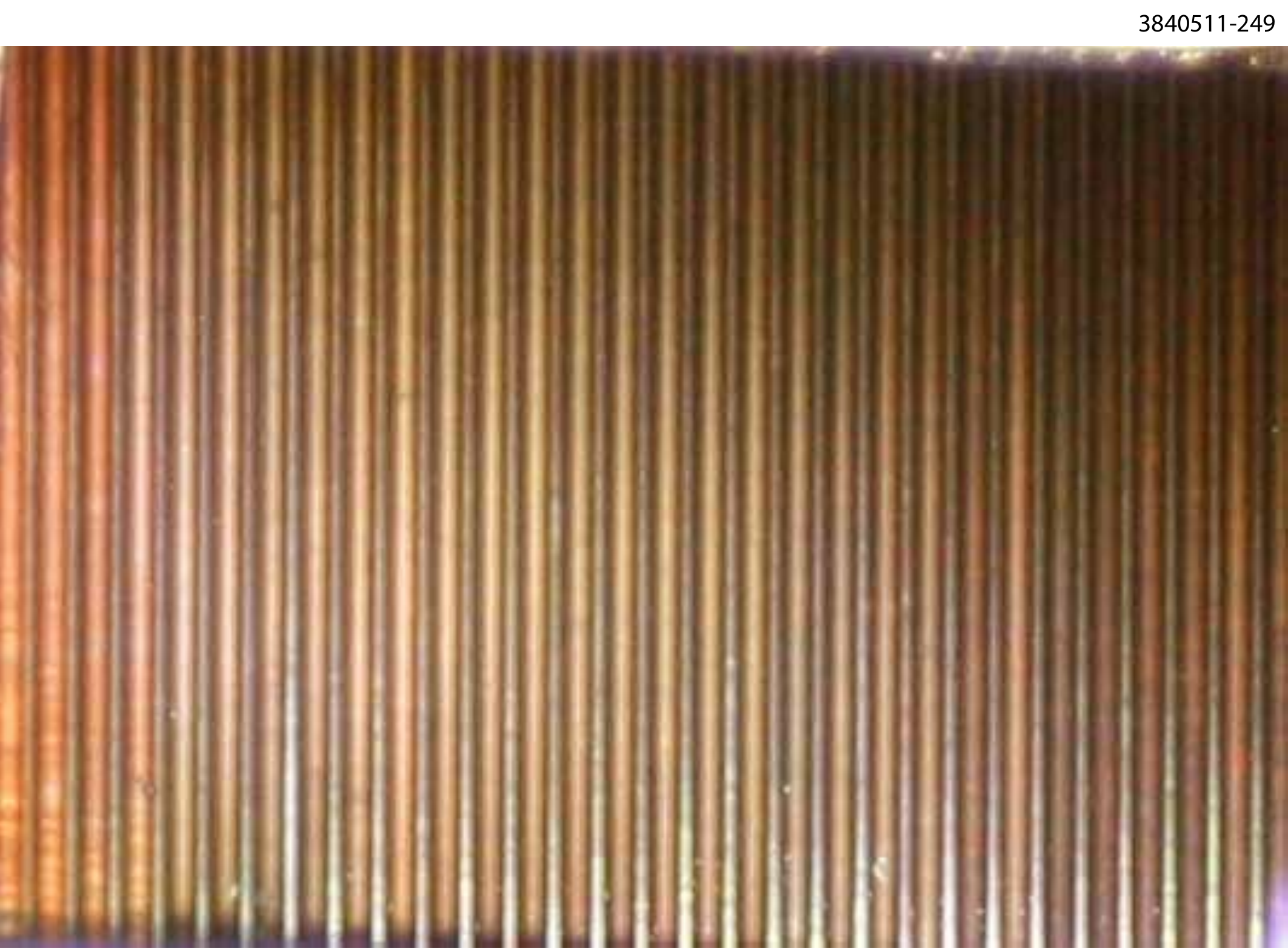} &
\includegraphics[width=0.45\textwidth]{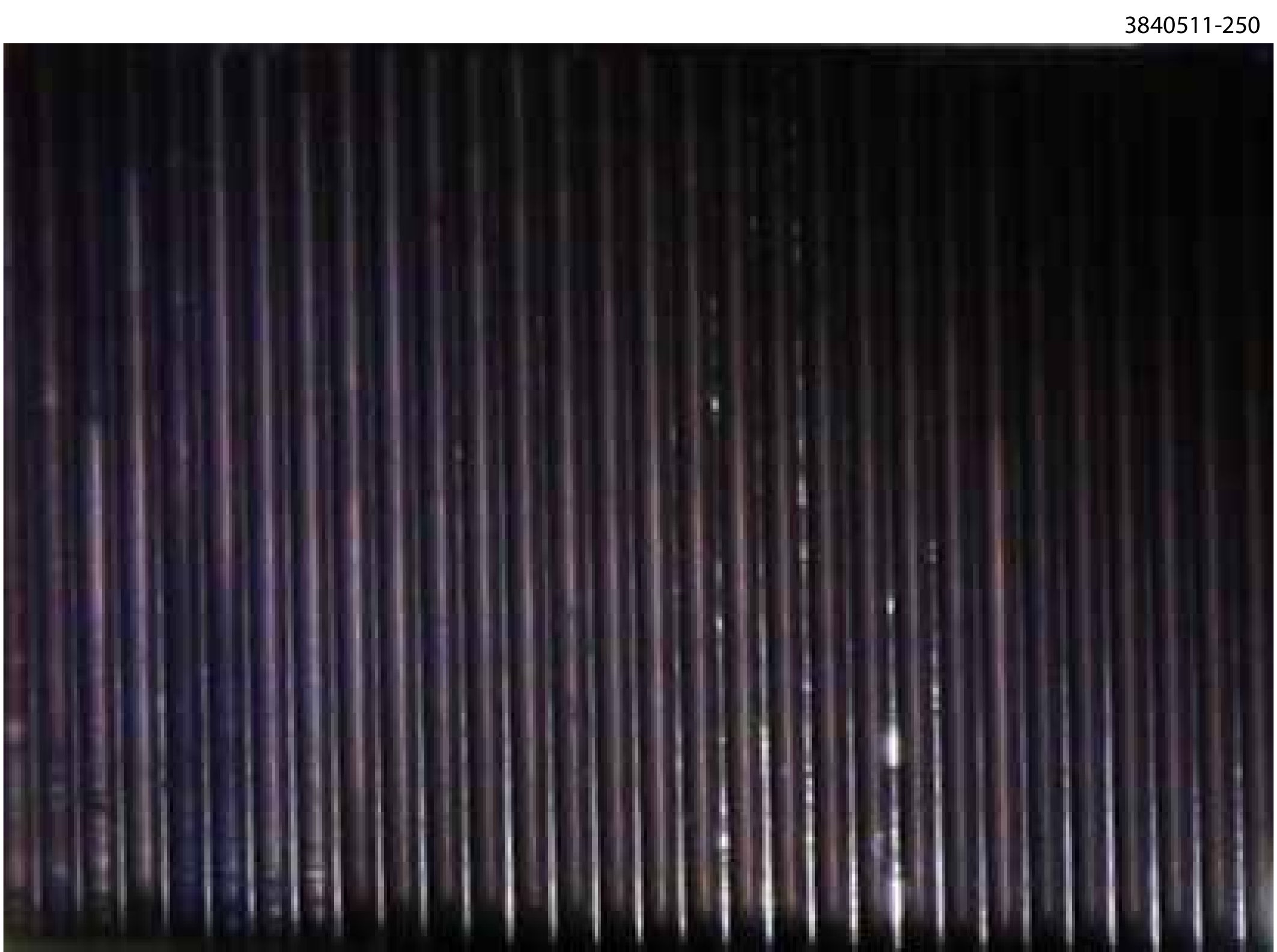}\\
\end{tabular}
	\caption[SCW RFA beam pipe Groove Inspection Photos]{The grooves of the SCW RFA beam pipe were inspected optically.  LEFT: Before TiN coating; RIGHT:After TiN coating.  \label{fig:SCW_groove_inspection}}
\end{figure}

\begin{itemize}
	\item During the CHESS runs (and some of the {\cesrta} experimental runs), short electron/positron bunches with high bunch current were stored. Therefore, concerns were raised as to whether the high image current density on the sharp groove tips may have caused over-heating that resulted in damage. Close-up optical inspection of the grooves was carried out with a combination of a specially design 'trolley' (that locates a mirror above the grooves in the beam pipe) and a high-zoom digital camera.  The entire length of the groove plate was inspected for damage to the groove tips. No observable damage was found.
	\item Using a DC sputtering technique, a TiN thin film coating was applied to the interior surfaces of the beam pipe, including the grooved portion at the bottom beam pipe.  The coating setup is shown in Figure~\ref{fig:SCW_groove_TiN_setup}.  Figure~\ref{fig:SCW_TiN_glow} displays the uniform discharge during the TiN thin film deposition.  The TiN deposition was undertaken in two stages with the titanium cathodes shifted transversely to minimize the geometric shadowing of the steep grooves.  The estimated TiN film thickness was 150 to 500~nm, based on the on-line measurements of a QCM (Quartz Crystal Micro-balance) and the off-line measurements of witness coupons.
	\item Close-up optical inspection of the grooves was done after the TiN coating.  Figure~\ref{fig:SCW_groove_inspection} shows typical inspection images of a portion of the grooves before and after the TiN coating.  Although the camera did not possess sufficient resolution for the fine features of the sharp tips/troughs, no obvious sign of over-heating and damage to the grooves were visible.  The image also showed relatively uniform TiN coating with no significant shadowing within the groove troughs.
\end{itemize}

	The TiN coating process was performed in the presence of RFA assemblies with all RFA grids and collectors grounded.  The RFAs were found to be fully operational after coating in this manner.  The TiN coated, grooved RFA SCW assembly was reinstalled in the L0 EC experimental region in SCW02WA location in January~2011. 

\subsection{Clearing Electrode Chamber}

\begin{figure}
    \centering
    \includegraphics[width=0.7\columnwidth, angle=-90, width=6.0in]{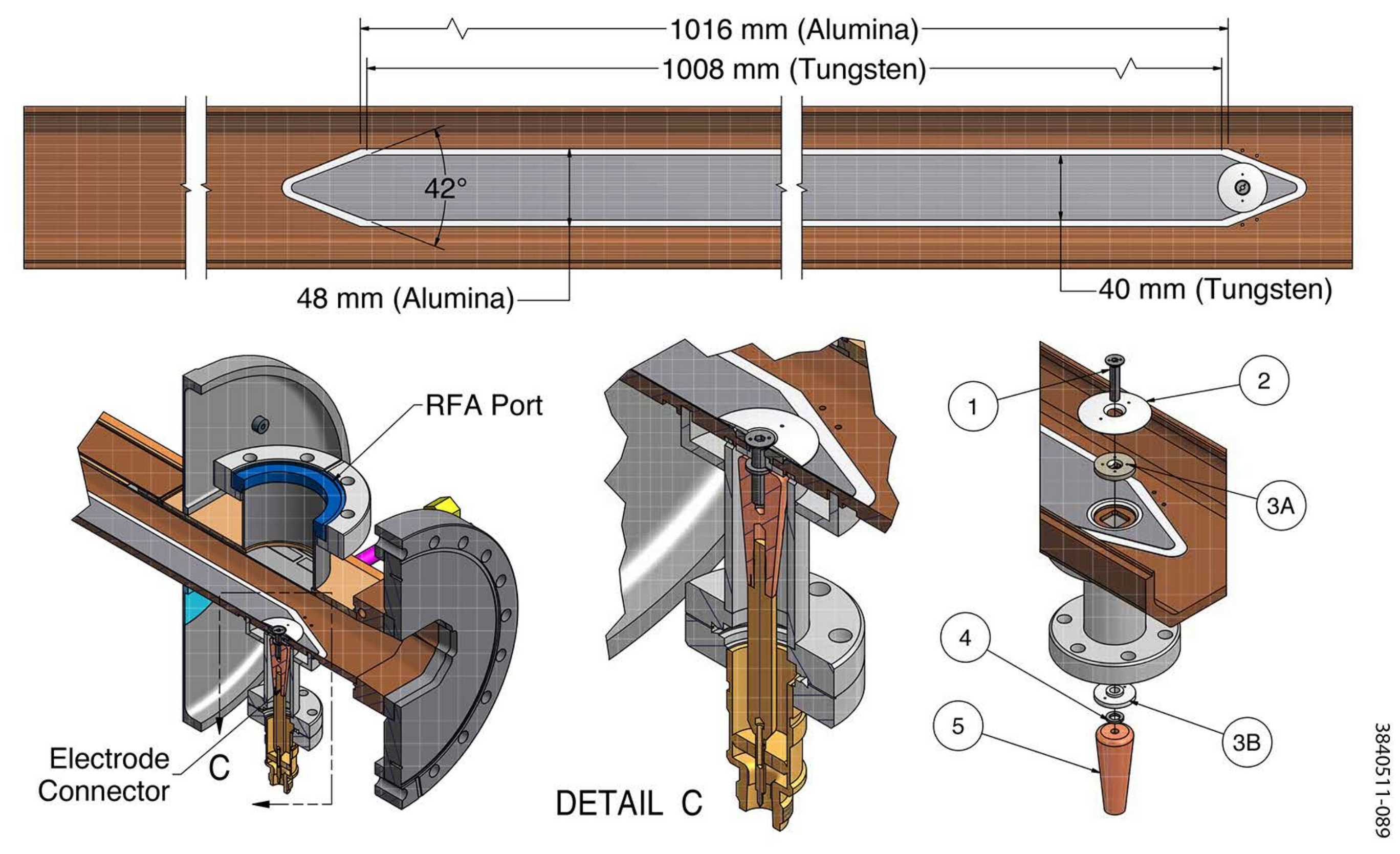}
    \caption[SCW RFA beam pipe with a bottom groove plate with clearing electrode assembly.]{\label{fig:SCW_RFA_pipe_electrode}
	SCW RFA beam pipe with clearing electrode assembly.  Top: sectional view and clearing electrode as seen from the top. Bottom left: sectional view  of the clearing electrode and the HV input port. Bottom middle: detail of HV coupling port. Bottom right: exploded view of HV port;  electrical contact is made with a vented screw (1), an aluminum alloy button (2), insulation spacer made of polyether ether ketone (PEEK) (3A, 3B),  a stainless steel spring washer (4), and a tapered copper pin (5).}
\end{figure}

One of the electron cloud mitigation methods studied at {\cesrta} makes use of a clearing electrode, operating with a positive bias voltage to attract the electrons, which would ordinarily collect around the positron beam.  To investigate its EC suppression efficacy in the wiggler field, a clearing electrode was installed on the bottom of an RFA-equipped beam pipe in a SCW.  The structure of the beam pipe is shown in Figure~\ref{fig:SCW_RFA_pipe_electrode}.  The beam pipe consists of a top half with three RFAs (identical to the design as shown in Figure~\ref{fig:SCW_RFA_structure}) and a bottom half with a clearing electrode assembly. 

\begin{figure}
    \centering
    \includegraphics[width=0.75\columnwidth]{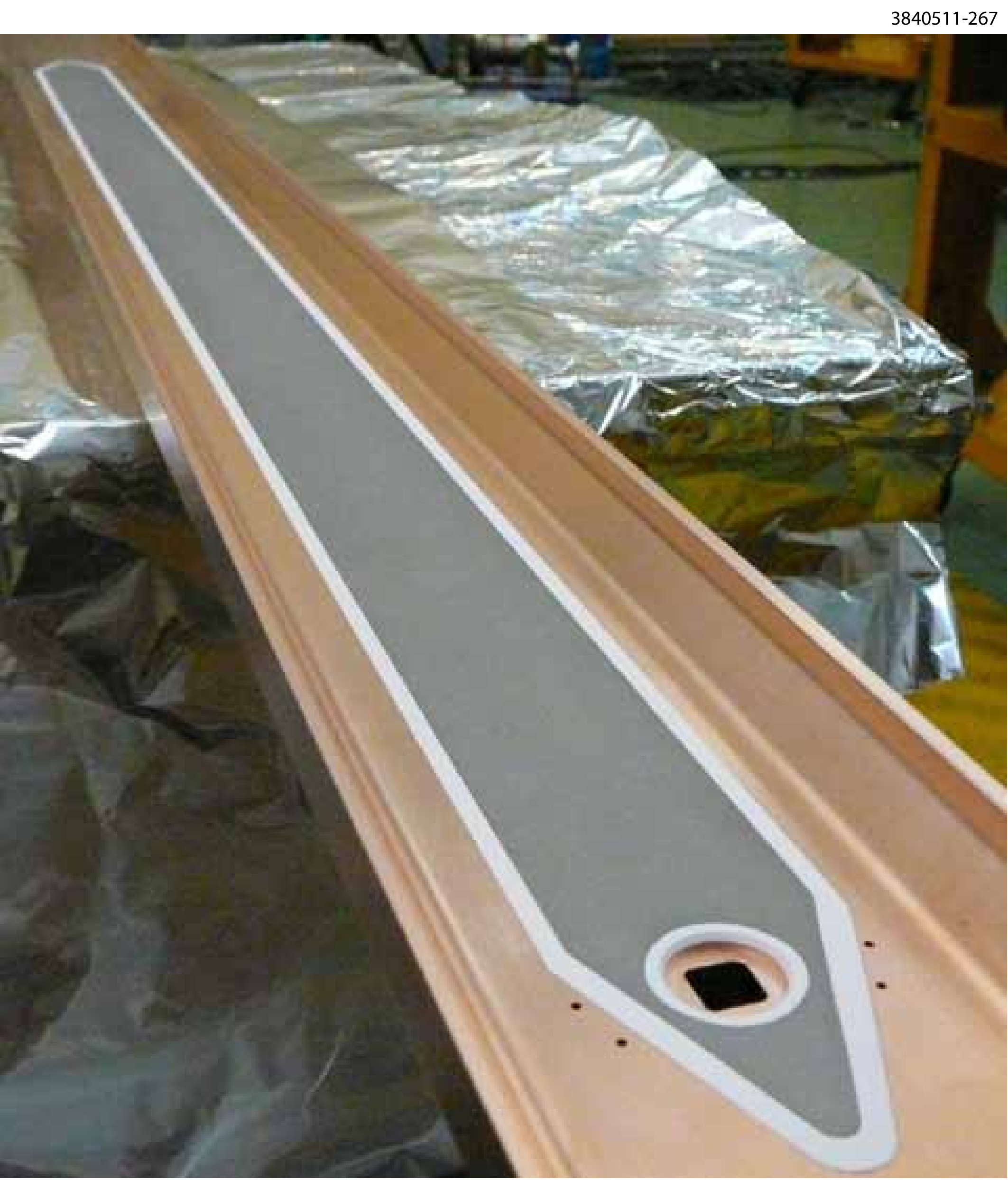}
    \caption[Photo of deposited electrode on bottom of the SCW beam pipe.]{\label{fig:SCW_electrode_deposited} Photo of deposited electrode on the bottom of the SCW beam pipe.}
\end{figure}

The geometry of the clearing electrode is a thin stripline \cite{NIMA598:372to378}, which consists of an alumina ceramic layer on copper surface (of approx. 0.2~mm in thickness) as an insulator and a thin layer of tungsten on the ceramic surface (of approx. 0.1~mm in thickness) as the electrode.  The area of the alumina ceramic layer was larger than that of the electrode, so that the required DC voltage could be applied to the electrode in vacuum.  These two layers were deposited via a thermal spray technique and were tightly bonded to the copper chamber.  Tungsten was chosen as the electrode material owing to its small thermal expansion rate (as compared to the alumina ceramics), good thermal and electrical properties and substantial experience in  thermal spray applications.  The photograph in Figure~\ref{fig:SCW_electrode_deposited} shows the finished bottom beam pipe with its deposited electrode.  The insulation resistance between the electrode and the copper chamber was about 5~M$\Omega$ (in dry air) and the electrode is capable of withholding DC voltages above 1~kV.

\begin{figure}
    \centering
    \includegraphics[width=0.75\columnwidth]{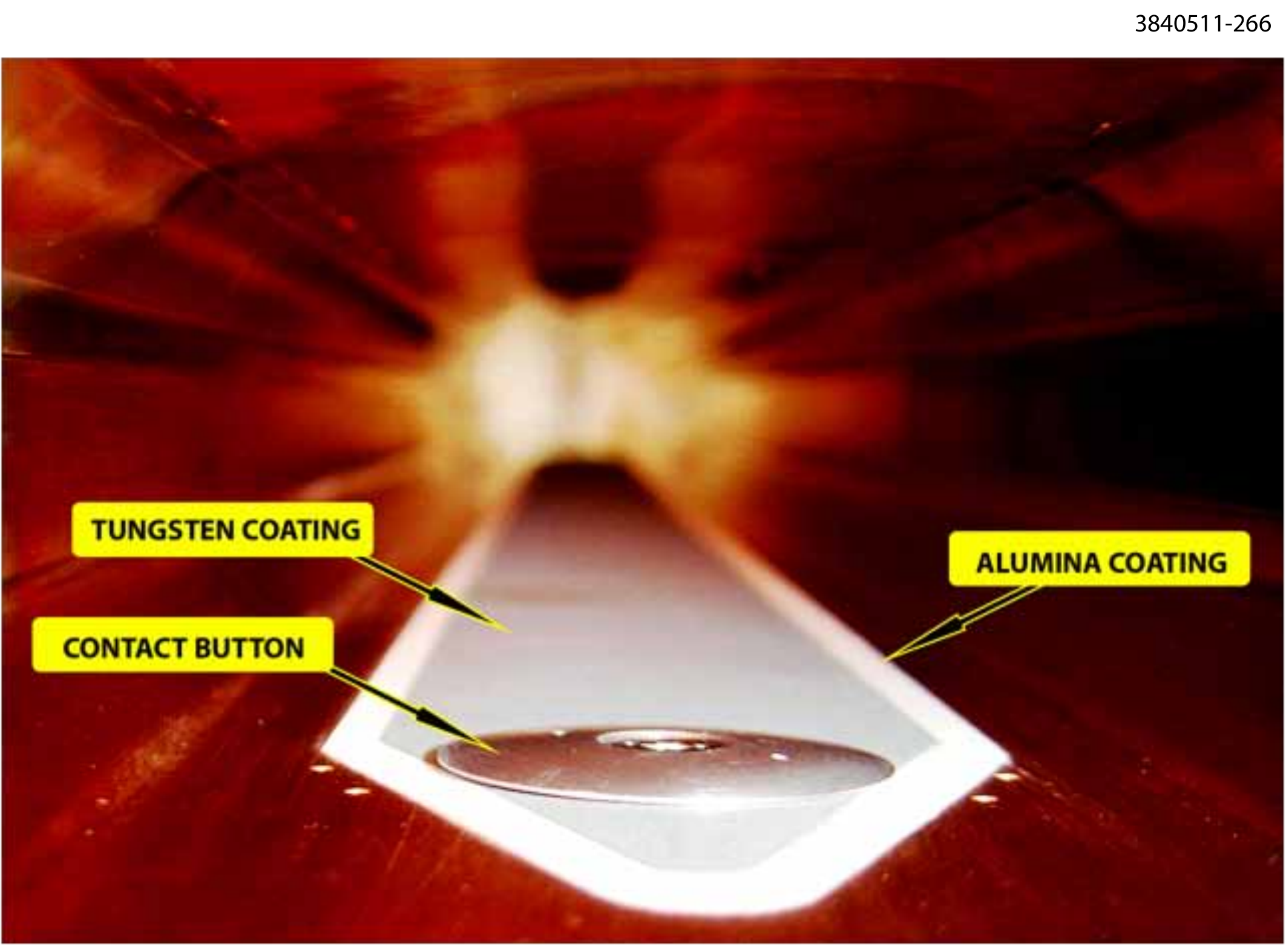}
    \caption[Photo of electrode connection button on the bottom of the SCW beam pipe.]{\label{fig:SCW_electrode_button} Photo of electrode connection button on the bottom of the SCW beam pipe.}
\end{figure}

Since the magnetic field within the wiggler magnet is vertical, electrons will stream along field lines, impacting the top and bottom walls of the vacuum chamber and will produce secondary electrons, which subsequently stream to the opposite vacuum chamber wall. Thus the clearing electrode is designed to have a width of 40~mm and is placed on the lower wall of the vacuum chamber, extending for 1.09~m along the beamline, in order to intercept electrons impacting the bottom of the vacuum chamber.  However, from  the standpoint of the beam-impedance, the narrower the electrode is, the smaller the impedance.  To minimize HOML induced by the electrode, the ends of the electrode are tapered at a 42$^\circ$ angle down to a 3~mm radius at its tip.  
The HV connection is made by a coaxial line coming through a port located underneath one of the tapered ends of the electrode.  The electrical connection to the electrode is made via a convex button washer (having a 26~mm diameter) on the top of the electrode, as shown in Figure~\ref{fig:SCW_RFA_pipe_electrode} and Figure~\ref{fig:SCW_electrode_button}.  The connection was designed to make the inner surface (carrying the image currents from the beam) as smooth as possible, while keeping a secure electrical contact.  Although the effects of the low profile of the discontinuity, the tapered ends of the electrode and the hidden HV connection have not been calculated in detail, they are expected to produce a HOML parameter of less than a few times 0.001~V/pC for the 7 to 10~mm bunch length of CESR.  Any heating from the wall currents flowing on the surface of the electrode is easily handled by conduction to the beam pipe through the thin alumina dielectric layer, which has good thermal conductivity.  Thermocouples mounted on the bottom beam pipe near the electrode assembly have detected no increase in heating due to HOML at any of the beam currents during operations.

At LBNL the bottom half of the beam pipe (containing the thermal-sprayed electrode) was electron beam-welded with the top half beam pipe (with all RFA features as shown in Figure~\ref{fig:SCW_RFA_structure}) to form the pre-RFA beam pipe assembly and then delivered to Cornell in February~2010.  Following the same assembly procedures described earlier, RFAs were inserted into the electrode beam pipe, integrated into an SCW and then installed in the L0 experimental region for {\cesrta} experimental runs beginning April~2010.    

During stored beam operation one concern with the electrode design was the peak voltage induced by a bunch's passing.  There are two parts to this particular concern.  First is that the peak voltage is high enough to either break down the dielectric layer or produce an electrical discharge in the vacuum, which could cause metal to be plated onto the dielectric.  Either of these will degrade the high voltage standoff capability of the electrode.   The second concern is that the voltage induced onto the electrode might couple to subsequent bunches causing enhanced HOML or transverse kicks possibly destabilizing later bunches in the train.  This could be particularly a concern for the clearing electrode, since its structure resembles that of a stripline pickup or stripline kicker.  Since the transmission line impedance of the electrode structure is difficult to match to the external coupling port, there can be many reflections of the induced voltage before it damps out.  If the transmission line-electrical length of the electrode is chosen poorly, a resonant excitation of the beam-induced voltage could result in large voltages occurring on the electrode with respect to ground.  This resonant enhancement of the voltage could lead to a sizable multiplication of all effects.  Therefore the 1.05~m length of the clearing electrode was designed, after taking into account the dielectric constant of alumina of 9.8 (a reduction of the propagation delay of the signal in the stripline by a factor of 3.13 from the speed of light in vacuum), to have a round trip delay of 21.9~ns, which is not a multiple of either 4~ns or 14~ns, the primary bunch spacings employed for {\cesrta}.   

To reduce the peak induced voltage, the ends of the electrode are tapered.  As a bunch passes by the tapered end of the electrode the wall currents induced in the electrode spread out over a time equal to the difference of the propagation delay of the signal in the electrode's stripline transmission mode and the transit time of the bunch.  With a taper length of 31.3~mm, the duration of the induced signal at both ends will spread out over 0.2~ns in the forward transit direction and 0.4~ns in the reverse direction.  Since the temporal standard deviation of the bunch ranges between 20 and 35~ps, spreading the induction of the signal in the electrode over more than ten times the bunch's rise time will reduce the peak signal by a comparable factor.  Since the wall currents traveling with the beam are induced into the transmission line composed of the electrode resting on the dielectric layer on top of the vacuum chamber wall, the induced voltage will be proportional to the impedance of the transmission line.  With a dielectric thickness of 0.2~mm and a maximum electrode width of 44~mm, the transmission line impedance is 0.59~$\Omega$.  At the tapered ends the impedance increases to 3.6~$\Omega$, thus the reflections off of the ends will cause additional (somewhat more complicated) spreading of the transmission line signal.  

Some care was taken to make a reasonable match of the impedance of the external coaxial HV connection to the clearing electrode.  The effective impedance of the electrode at the point where the external port attaches to the electrode's taper is 0.78~$\Omega$.  The hole through the beam pipe under the taper plus the HV connection post (sticking through the hole) have an impedance of 6.2~$\Omega$, making it impossible to match the clearing electrode's impedance at the connection point.  However, some improvement to the mismatch to the 50~$\Omega$ type-N connector was accomplished by tapering the center conductor with a conical taper from the 50~$\Omega$ connector to 11.5~$\Omega$ where the center conductor attaches to the HV connection post.  So in the final design there are two mismatches between electrode and the type-N connector: 1) between the electrode and the HV post and 2) between the HV post and the tapered coaxial feed line, giving reflection coefficients, respectively, of $-0.80$ and $-0.30$.  To allow for thermal expansion, there is a sliding contact where the conical coaxial center conductor attaches to the constant diameter, 50~$\Omega$ center conductor.

After the construction and assembly of the electrode and the coaxial HV connection port, time domain reflectometer (TDR) measurements were made.  The TDR allows a basic check of the impedance of the entire structure using a step function-signal with a rise-time of 25~ps.  However, since the electrode's transmission line impedance is so low, only an upper limit may be placed on its impedance.  At the input side, the type-N connector had a reflection of +~6$\%$, while the sliding contact had additional $\pm~4\%$ reflections.  Neither of these is very serious as they are relatively small and are contained within regions of less than 0.15~ns propagation delays.  The estimated upper limit of the transmission line impedance of the central section of the clearing electrode itself is 2.6~$\Omega$, which is consistent with the expected 0.59~$\Omega$.

\begin{figure}[htbp] 
   \centering
   \includegraphics[width=0.6\columnwidth]{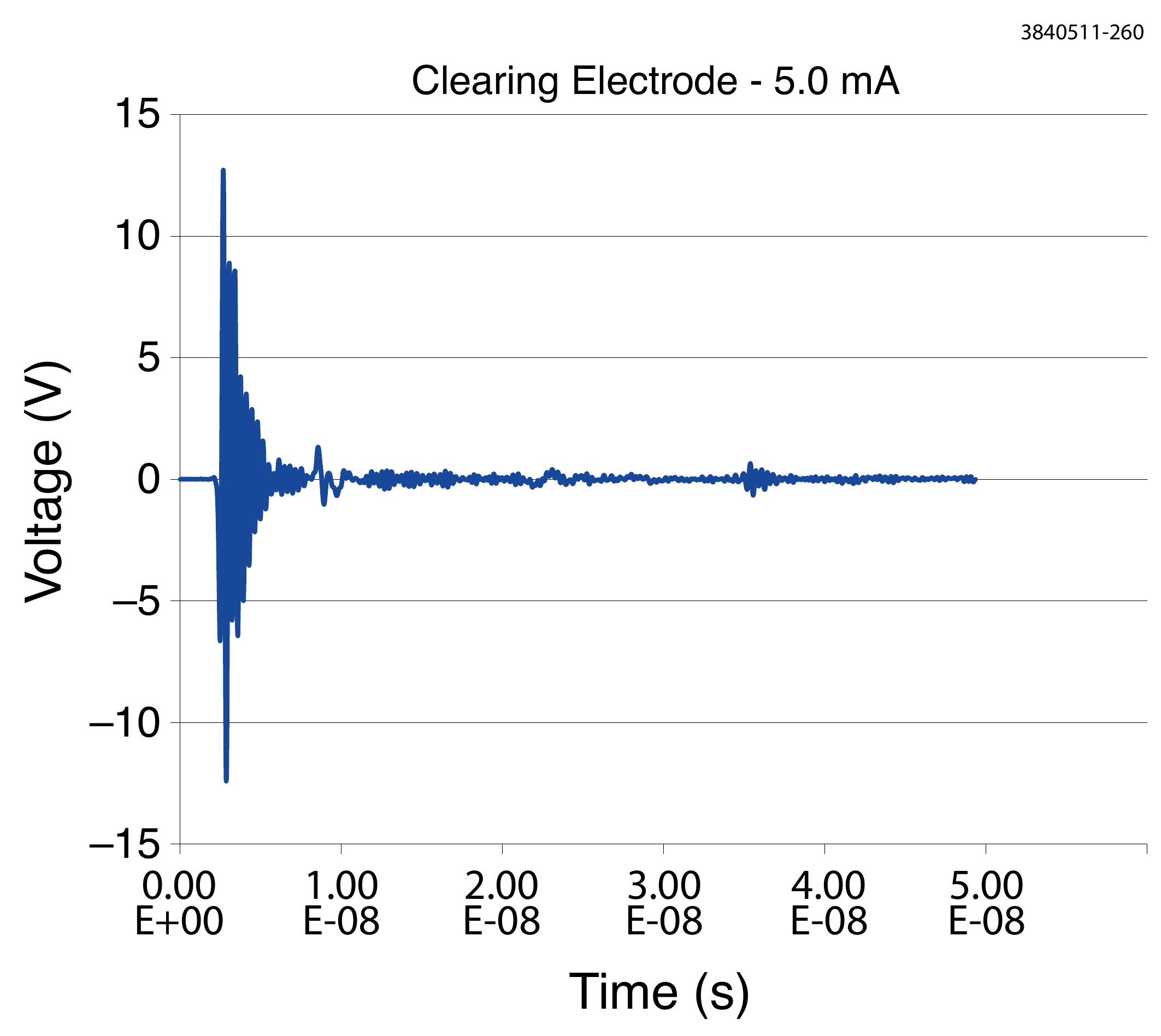} 
   \caption[Clearing electrode HV port signal for a single 5 mA positron bunch]{\label{fig:ClearingElectrode00008-LongView} 
   Clearing electrode HV port signal for a single 5 mA positron bunch}
\end{figure}

\begin{figure}[htbp] 
   \centering
   \includegraphics[width=0.6\columnwidth]{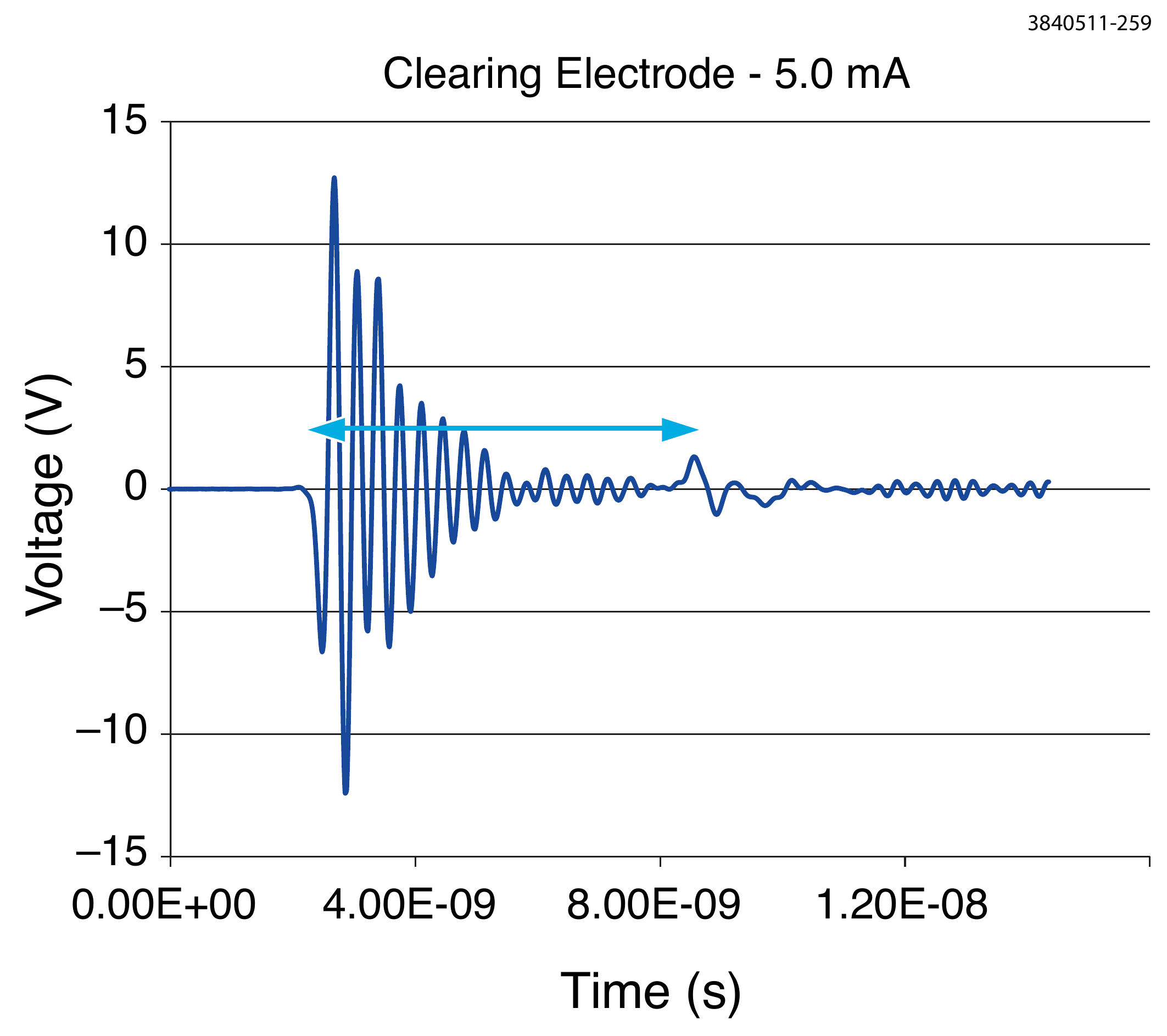} 
   \caption[Clearing electrode HV port signal for a single 5 mA positron bunch on an expanded time scale.]{\label{fig:ClearingElectrode00008-Closeup} 
   Clearing electrode HV port signal for a single 5 mA positron bunch on an expanded time scale.  The time segment marked with the arrows represents the delay between signals induced at the HV coupling port end and the far end of the electrode.}
\end{figure}

\begin{figure}[htbp] 
   \centering
   \includegraphics[width=0.6\columnwidth]{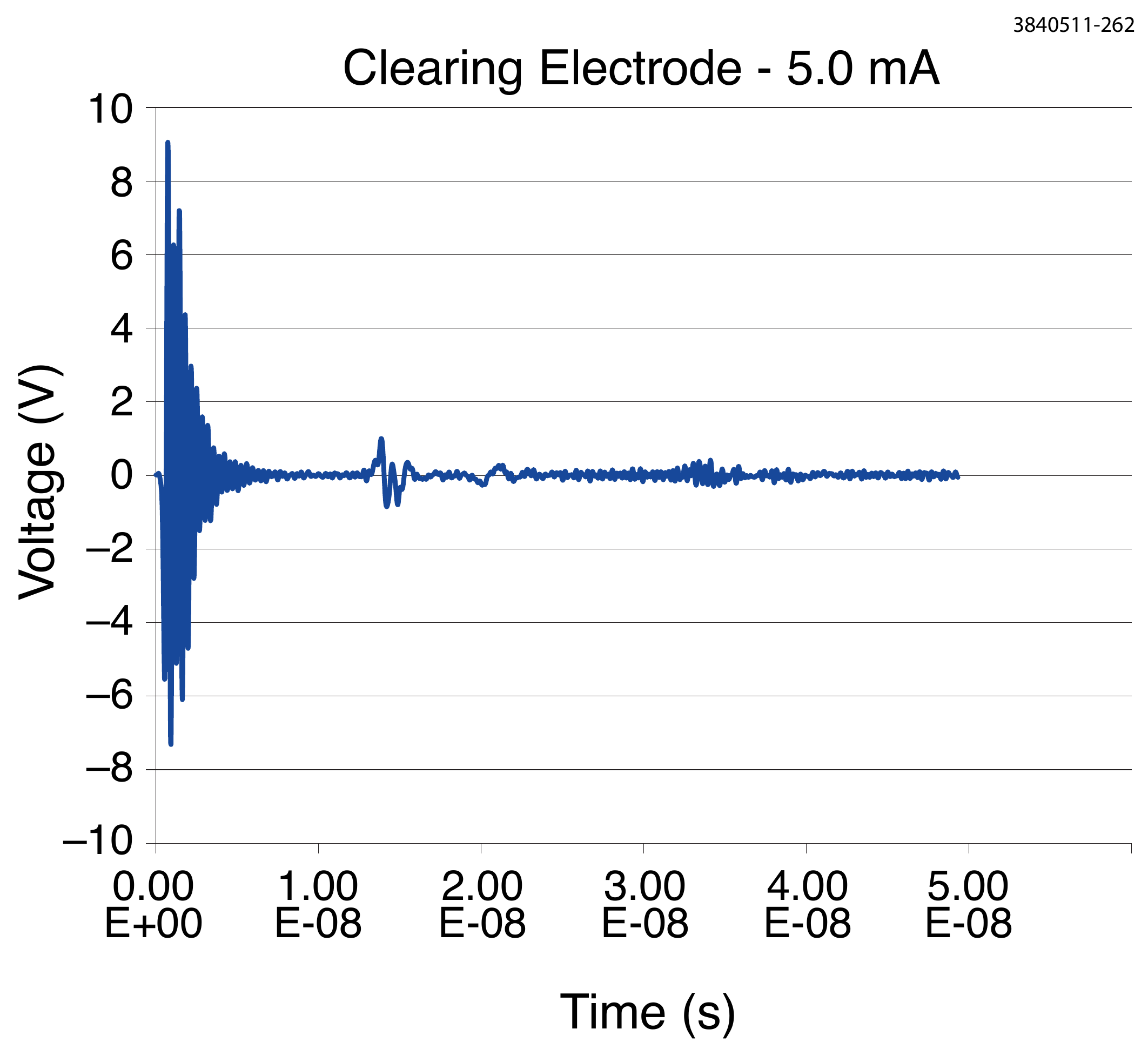} 
   \caption[Clearing electrode HV port signal for a single 5 mA electron bunch.]{\label{fig:ClearingElectrode00012-LongView} 
   Clearing electrode HV port signal for a single 5 mA electron bunch. }
\end{figure}

\begin{figure}[htbp] 
   \centering
   \includegraphics[width=0.6\columnwidth]{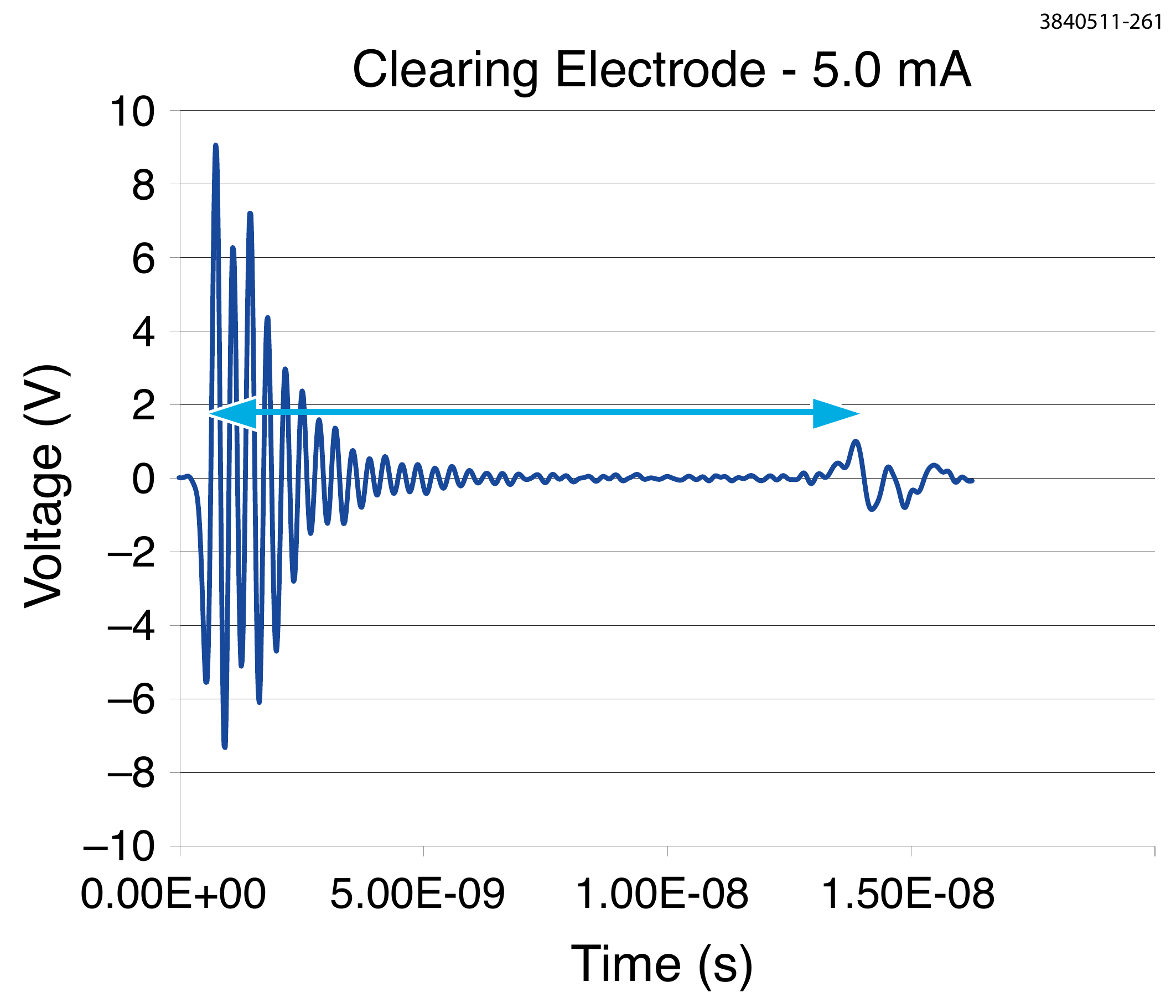} 
   \caption[ Clearing electrode HV port signal for a single 5 mA electron bunch on an expanded time scale.]{\label{fig:ClearingElectrode00012-Closeup} 
   Clearing electrode HV port signal for a single 5 mA electron bunch on an expanded time scale.  The time segment marked with the arrows represents the delay between signals induced at the HV coupling port end and the far end of the electrode.}
\end{figure}

With 5~mA (8.0x10\textsuperscript{10} particles) single bunch stored electron and positron beams, measurements were made to observe the signal coming from the type-N connector at the HV input port.  After installation in CESR the orientation of the clearing electrode has the HV input port oriented so that the electron bunches pass the end with the input port first and then the end with an open connection afterwards.  A positron bunch circulates in the opposite direction as compared with an electron bunch.   All measurements were made with 26~dB of attenuation for signal before connecting to a LeCroy Model Wavemaster~804ZI, 4~GHz bandwidth oscilloscope.  In the following figures the scales have been corrected for the 26~dB of attenuation.  Figures~\ref{fig:ClearingElectrode00008-LongView} and~\ref{fig:ClearingElectrode00008-Closeup} show the signal from the positron single bunch on longer and shorter time scales, respectively.  The signal from the positron bunch has a peak value of 13~V at the time of the bunch's passage by the HV coupling port as is indicated by the 2.9~GHz burst.  Although a detailed model for the coupling from this tapered stripline has not been constructed, this frequency appears to be characteristic of the propagation delay along the taper.  Figure~\ref{fig:ClearingElectrode00008-Closeup} shows in more detail the signal induced at the HV coupling port (as described above) and the open taper at the far end.  Although the signal from the positron bunch was induced at the open taper earlier, due to the propagation delay from the dielectric insulator, this signal arrives at the HV coupling port 0.61~ns later than the signal induced in the tapered end connected to the HV coupling port.  The time segment marked in figure~\ref{fig:ClearingElectrode00008-Closeup} is approximately the delay from the $v=c$ beam transit velocity and the $v=0.31~c$ signal propagation velocity.  Likewise, the signal observed from the 5~mA electron bunch, seen in figures~\ref{fig:ClearingElectrode00012-LongView} and ~\ref{fig:ClearingElectrode00012-Closeup} (viewed on longer and shorter timescales, respectively), shows the same high frequency burst from the taper at the HV coupling port followed by the lower bandwidth pulse from the open tapered end of the electrode.  In this case the peak signal observed in the HV input port is 9~V and the delay between the signals at the two ends has increased to 1.33~ns.  The longer delay is caused by the fact that the bunch must transit the length of the electrode at $v = c$ and then the signal induced in the open taper must propagate back at $v = 0.31~c$.  By comparing the delays between these two signals for positrons and electrons, it is possible to determine the measured propagation delay in the transmission line defined by the clearing electrode and, hence, the dielectric constant for the alumina.  These measurements yield a dielectric constant of 2.9 in good agreement with the expected 3.13 value.

Neither of the positron or electron beam induced signals, as viewed at the HV coupling port, suggest serious peak voltages for the electrode at any typical operating currents for {\cesrta}.  They also suggest that the reflections decay in only a small number of traversals along the electrode and that taken with the measured propagation delays imply that resonant build-up of the signals from multiple bunches within trains will not be a problem.

\begin{figure}[htbp] 
   \centering
   \includegraphics[width=0.6\columnwidth]{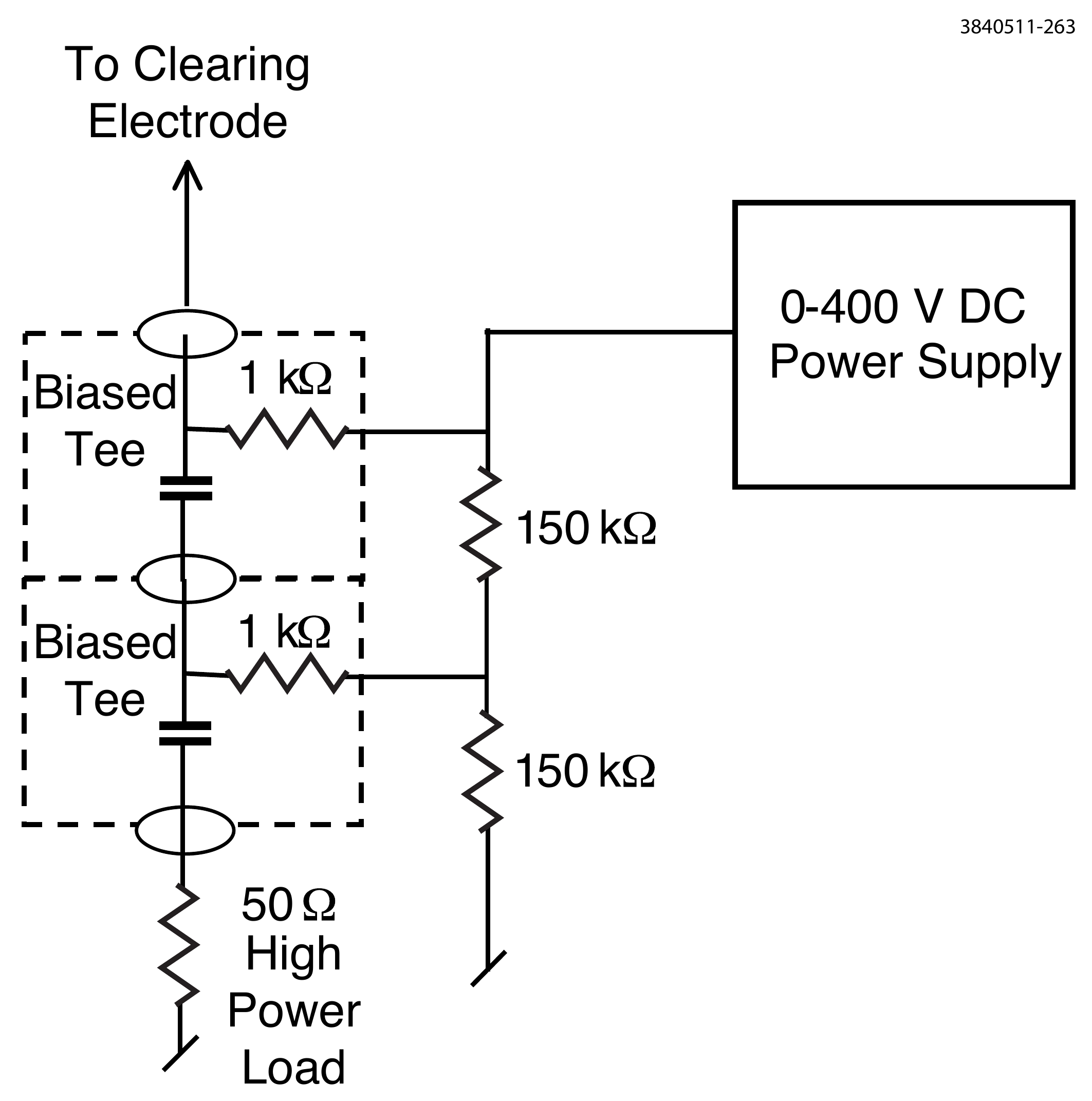} 
   \caption[Coupling network for the clearing electrode's HV coupling port .]{\label{fig:ClearingElectrodeCouplingNetwork} 
   Coupling network for the clearing electrode's HV coupling port.}
\end{figure}

\begin{figure}[htbp] 
   \centering
   \includegraphics[width=0.6\columnwidth]{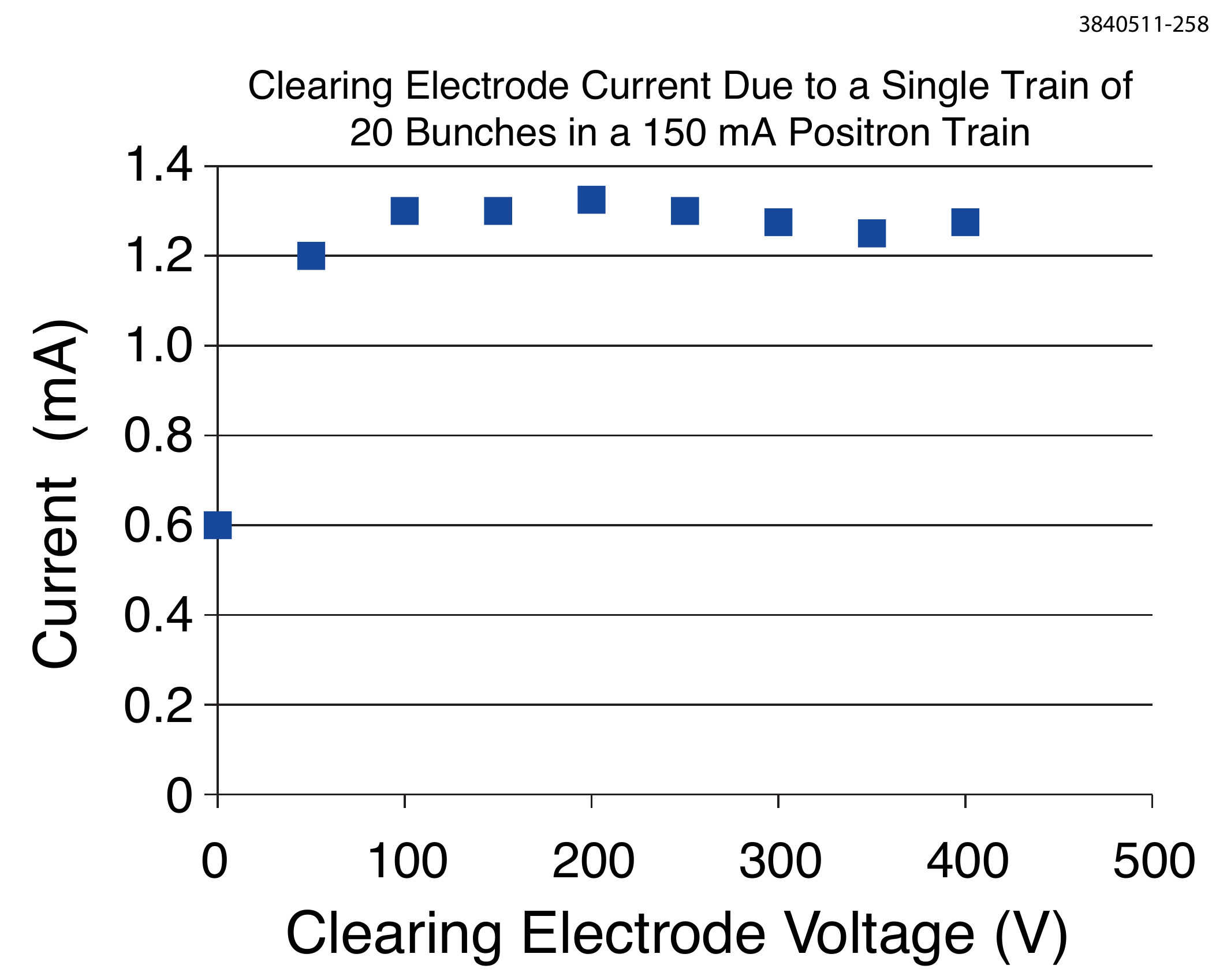} 
   \caption[Beam induced clearing electrode current as a function HV bias setting.]{\label{fig:Clearing_Electrode_Currents} 
   Beam induced clearing electrode current as a function HV bias setting for a 20 bunch train of a total current of 150 mA at 5.3 GeV with the wiggler turned off. }
\end{figure}

The last of the post-installation measurements was made after attaching the HV coupling network and measuring the electrode current vs. electrode voltage for a given beam current.  The coupling network and its HV DC power supply are shown in figure~\ref{fig:ClearingElectrodeCouplingNetwork}; the network is constructed from bias tees, which were purchased from MECA (part number 200N-MF-1), capable of a peak 200~V~DC bias and rated to withstand a 300~W average RF power and 3~kW peak RF power.  Two of these were cascaded to obtain the needed 400~V bias capability for the {\cesrta} studies.  During bench testing of the MECA units, it was apparent that they had an unpleasant notched response just below about 100~MHz.  Anticipating that this could be a problem, the bias tees that were modified by replacing the originally installed inductor with a 1~k$\Omega$ resistor.  The frequency response is improved and now has a single pole high pass cutoff frequency of about 30~MHz.  Figure~\ref{fig:Clearing_Electrode_Currents} gives the measured clearing electrode current as a function of bias voltage for a train of 20 positron bunches at 5.3~GeV, while the superconducting wiggler is unpowered.   Note that 100~V of bias is enough to saturate the clearing electrode's collection current.

After operations for three {\cesrta} runs and two CHESS X-ray User runs (from April to December~2010) with peak bunch currents of over 9~mA, a visual inspection of the electrode and the electric contact was performed during the January~2011 shutdown.   The clearing electrode's appearance was the same as it was during its installation.  With an accumulated total beam dose exceeding 1000~Amp$\cdot$hr over the {\cesrta} and CHESS runs, the electrode and the electrical contact were found to be in excellent condition with no sign of any arcing or plating of material anywhere inside of the vacuum chamber. 


\subsection{Experimental Results}

Over many \cesrta~experimental runs data was acquired using the RFAs installed in the SCWs.  
This included studies in each of the experimental chambers as described above.  Results of some of these
measurements may be found in the following publication\cite{NIMA770:141to154}.

%% file: CesrConvWiggler-JINST.bbl
\providecommand{\bysame}{\leavevmode\hbox to3em{\hrulefill}\thinspace}
\providecommand{\MR}{\relax\ifhmode\unskip\space\fi MR }
\providecommand{\MRhref}[2]{%
  \href{http://www.ams.org/mathscinet-getitem?mr=#1}{#2}
}
\providecommand{\href}[2]{#2}
\begin{thebibliography}{10}

\bibitem{JINST10:P07012}
M.~Billing, \emph{{T}he conversion of {CESR} to operate as the {T}est
  {A}ccelerator, {C}esr{TA}. {P}art 1: overview}, J. Instrum. \textbf{10}
  (2015).

\bibitem{JINST10:P07013}
M.~G. Billing and Y.~Li, \emph{The conversion of {CESR} to operate as the test
  accelerator, {CesrTA}, part 2: Vacuum modifications}, J. Instrum. \textbf{10}
  (2015).

\bibitem{JINST11:P04025}
M.~G. Billing, J.~V. Conway, J.~A. Crittenden, S.~Greenwald, Y.~Li, R.~E.
  Meller, C.~R. Strohman, J.~P. Sikora, J.~R. Calvey, and M.~A. Palmer,
  \emph{The conversion of {CESR} to operate as the test accelerator, {CesrTA},
  part 3: Electron cloud diagnostics}, J. Instrum. \textbf{11} (2016).

\bibitem{PAC03:WPPE037}
Y.~He, G.~Codner, R.~D. Ehrlich, Y.~Li, V.~Medjidzade, A.~Mikhailichenko, N.~B.
  Mistry, E.~Nordberg, D.~Rice, D.~Sabol, K.~Smolenski, D.~Widger, and E.~N.
  Smith, \emph{Design and operation of the cryostat for the {CESR-c}
  superconducting wiggler magnets}, Proceedings of the 2003 Particle
  Accelerator Conference, Portland, OR (Joe Chew, Peter Lucas, and Sara Webber,
  eds.), IEEE, 2003, pp.~2399--2401.

\bibitem{IPAC12:TUPPR065}
J.~A. Crittenden, M.~A. Palmer, and D.~L. Rubin, \emph{Wiggler magnet design
  development for the {ILC} damping rings}, Proceedings of the 2012
  International Particle Accelerator Conference, New Orleans, LA, IEEE, 2012,
  pp.~1969--1971.

\bibitem{PAC07:TUPAS067}
L.~Wang and F.~Zimmermann, \emph{Electron cloud in the wigglers of the positron
  damping ring of the {I}nternational {L}inear {C}ollider}, Proceedings of the
  2007 Particle Accelerator Conference, Albuquerque, NM (C.~Petit-Jean-Genaz,
  ed.), IEEE, 2007, pp.~1808--1810.

\bibitem{NIMA760:86to97}
J.~R. Calvey, W.~Hartung, Y.~Li, J.~A. Livezey, J.~Makita, M.~A. Palmer, and
  D.~Rubin, \emph{Comparison of electron cloud mitigating coatings using
  retarding field analyzers}, Nucl. Instrum. Methods Phys. Res. \textbf{A760}
  (2014), 86--97.

\bibitem{ECLOUD04:139to141}
G.~Stupakov and M.~Pivi, \emph{Suppression of the effective secondary emission
  yield for a grooved metal surface}, Proceedings of ECLOUD 2004: 31st ICFA
  Advanced Beam Dynamics Workshop on Electron-Cloud Effects, Napa, CA (Geneva,
  Switzerland) (M.~Furman, S.~Henderson, and F.~Zimmerman, eds.), no.
  CERN-2005-001, CERN, 2004, pp.~139--141.

\bibitem{JAP104:104904}
M.~Pivi, F.~K. King, R.~E. Kirby, T.~O. Raubenheimer, G.~Stupakov, and
  F.~Le~Pimpec, \emph{Sharp reduction of the secondary electron emission yield
  from grooved surfaces}, J. Appl. Phys. \textbf{104} (2008).

\bibitem{VACUUM73:195to199}
A.~A. Krasnov, \emph{Molecular pumping properties of the {LHC} arc beam pipe
  and effective secondary electron emission from {Cu} surface with artificial
  roughness}, Vacuum \textbf{73} (2004), 195--199.

\bibitem{NIMA571:588to598}
L.~Wang, T.~O. Raubenheimer, and G.~Stupakov, \emph{Suppression of secondary
  emission in a magnetic field using triangular and rectangular surfaces},
  Nucl. Instrum. Methods Phys. Res. \textbf{A571} (2007), 588--598.

\bibitem{LWang:PrivateC}
L.~Wang, private communication.

\bibitem{NIMA598:372to378}
Y.~Suetsugu, H.~Fukuma, L.~Wang, M.~Pivi, A.~Morishige, Y.~Suzuki,
  M.~Tsukamoto, and M.~Tsuchiya, \emph{Demonstration of electron clearing
  effect by means of a clearing electrode in high-intensity positron ring},
  Nucl. Instrum. Methods Phys. Res. \textbf{A598} (2009), 372--378.

\bibitem{NIMA770:141to154}
J.~R. Calvey, W.~Hartung, Y.~Li, J.~A. Livezey, J.~Makita, M.~A. Palmer, and
  D.~Rubin, \emph{Measurements of electron cloud growth and mitigation in
  dipole, quadrupole, and wiggler magnets}, Nucl. Instrum. Methods Phys. Res.
  \textbf{A770} (2015), 141--154.

\end{thebibliography}
